\documentclass[article,nofootinbib]{revtex4}

\usepackage{pdfpages}
\usepackage{epsf}
\usepackage{subfigure}
\usepackage{amsmath}
\usepackage{eqnarray,amsmath}
\usepackage{epstopdf}
\usepackage{mathrsfs}
\usepackage{natbib}
\usepackage{amssymb,latexsym}
\usepackage{dcolumn}
\usepackage{graphicx}
\usepackage{color}

\makeatletter
\newcommand*{\rom}[1]{\expandafter\@slowromancap\romannumeral #1@}
\makeatother
\def\be{\begin{equation}}
	\def\ee{\end{equation}}
\def\ba{\begin{eqnarray}}
	\def\ea{\end{eqnarray}}

\graphicspath{{images/}}
\begin{document}
	\title{\large \bf Ramsey Interferometers as a test for the correction to quantum mechanics  }

	\author{Abasalt Rostami} 
	\affiliation{Department of Physics, Sharif University of Technology, Tehran, Iran } 
	\affiliation{ School of Physics, Institute for Research in Fundamental Sciences (IPM), P. O. Box 19395-5531, Tehran, Iran } 
	\email{aba-rostami@ipm.ir}     
	
	\author{Javad T. Firouzjaee}
	\affiliation{ School of physics, Institute for Research in Fundamental Sciences (IPM), P. O. Box 19395-5531, Tehran, Iran } 
	\email{j.taghizadeh.f@ipm.ir}

		\author{Mehdi Golshani} 
			\affiliation{Department of Physics, Sharif University of Technology, Azadi Avenue, Tehran, Iran } 
			\affiliation{ School of physics, Institute for Research in Fundamental Sciences (IPM), P. O. Box 19395-5531, Tehran, Iran } 
	\email{golshani@sharif.edu}

    \begin{abstract}
        By applying the basic concept of the density matrix in an open quantum system and modification of quantum mechanics, we derive  Kossakowski-Lindblad equation and different properties of this equation are reviewed. Next, a pedagogical approach is used to present Ramsey's trick for linear modification of the quantum mechanics. We discuss how an open quantum mechanics or its modification changes the fraction of excited states in Ramsey Interferometers.

    \end{abstract}
    %
    %\pacs{}
    %
    \maketitle
    \tableofcontents
    
    \newpage
    
    %%%%%%%%%%%%%%%%%%%%%%%%%%%%%%%%%%%%%%%%%%%%%%%%%%
    \section{Introduction}
    %%%%%%%%%%%%%%%%%%%%%%%%%%%%%%%%%%%%%%%%%%%%%%%%%%
    
    When a measurement is done in quantum physics,  the wave-function of the quantum system collapses from a superposition of the eigenstates of the measured observable to one of the eigenstates. If one assumes the quantum superposition principle as a universal one, it would be possible to observe classical macro objects in a superposition of two distinct positions. Nevertheless, there are no signs of superposition states at the macro-scale so far. Classical physics governs over macro objects, while big manifestations of superposition states have been measured at the micro-scale. No superposition detection at macro-scale,  immediately raises these questions: Does the superposition principle really keep at the macro-scale physics? Can we distinguish a boundary between the micro and the macro world? Which quantity distinguishes the boundary between macro-scale and micro-scale? Can we have the Schroedinger equation, which is deterministic and linear, and does not predict the collapse? Moreover, the final result of a measurement is random and its probability is being given by the Born rule. How can we have a probabilistic result when the initial condition is identified exactly?\\

    All of these questions are addressed as the measurement problem of Quantum Mechanics. Niels Bohr first presented Copenhagen Interpretation of Quantum
    Mechanics in 1920. This interpretation states that the process of measurement gives the collapse of the wavefunction of the superposed quantum states.  This collapse is postulated to occur to the Born probability rule,     and no dynamical mechanism is specified to explain how the collapse occurs. Copenhagen Interpretation is in obvious disagreement with the linearity of the Schroedinger equation. In fact, this interpretation does not solve the quantum measurement problem,
    nor does it explain the absence of macroscopic superpositions. In this approach, the Bayesian statistical rule appears in the reduction of the state vector, by relating the density matrix after measurement to the density matrix before measurement \cite{Griffiths-book}\\

    Other solutions or interpretations  for the measurement problem can be classified into two categories: First, is the solutions which do     not change the implicit dynamics of quantum mechanics,
    such as decoherence \cite{decoherence}, the many-worlds interpretation
    \cite{Everett:1957hd} and Bohmian mechanics \cite{Bohm:1951xw}; Second those solutions
    which change the  dynamics of quantum
    mechanics, such as the spontaneous collapse dynamics \cite{Collapse-GRW,collapse-pearle,gravity-collapse,Bassi:2003gd,Bassi:2012bg}.
    
    Although quantum decoherence destroys interference among the states,  it cannot destroy superposition of states in the measurement problem, because it operates within the framework of linear quantum mechanics. In fact,
    decoherence causes quantum probability distributions to appear as classical probabilities. As a result, decoherence seems ineffective to explain the collapse of superposition for isolated macroscopic systems where there are no environmental degrees of freedom, such as the whole universe.
    
    The Many-worlds interpretation describes that Schroedinger evolution is universally valid at a different branch of universes, and the breakdown of superposition during a measurement is only apparent, not real. Many-worlds interpretation cannot explain the origin of probabilities and the Born probability rule, because it assumes that the evolution is deterministic throughout the measurement process.\\
    
%    \textbf{Bohmian Mechanics...Dr. Golshani may help us to complete this part}\\

    Collapse models \cite{Collapse-GRW,collapse-pearle,gravity-collapse,Bassi:2003gd,Bassi:2012bg} provide a well-defined phenomenology to solve the measurement problem. The first model for the collapse models was presented by Ghirardi, Rimini, Weber \cite{Collapse-GRW}, and then studied by Pearle \cite{collapse-pearle}. Those collapse models presume a universal stochastic noise that has a non-linear coupling with the matter. This non-linearity induces a localization in space, which destroys superpositions according to quantum probabilities. The strength of the coupling is fixed by phenomenological parameters defining the models, and those can have some experimental constraints (see \cite{Collapse-experimental} and references therein). 
    
    The collapse rate raises with the growth of the size and complexity of the system, and the effect of the collapse process becomes negligible at micro-scales, and is dominant when we go to macro-scales. In this mechanism, using a unique dynamical equation, both the quantum and the classical world can be described systematically. Since there is no justification from fundamental physical principles yet and there are different views about the physical origin of the collapsing field, these collapse models are phenomenological. However, some people have tried to find a natural explanation for collapse models by appealing to gravity, because gravity is universal and its strength increases with the mass of the system \cite{gravity-collapse}. \\

    Dynamical collapse models solve the quantum measurement problem by assuming that the Schroedinger equation is approximate.  It is an approximation to a stochastic nonlinear dynamics, and the stochastic nonlinear aspect becomes more important as one goes from microscopic scales to macroscopic ones. Since the collapse models must describe the primordial inhomogeneities in the cosmic microwave background radiation from a high inhomogeneous and isotropic state (Bunch-Davies vacuum) in the early universe, one has to develop a relativistic version for collapse models \cite{relativistic-collapse} which are consistent with general relativity.\\

    In this paper, we shall study the dynamics of an open system when the statistical properties of the system and its environment are independent of time. This study can also be used for examination of a general class of modified quantum theories. In this way, we shall have the Ramsey interferometer to find out if there are some observable differences between the standard theory of quantum mechanics and some modified versions. This study will be done in a detailed pedagogical way.\\
    
    This paper is organized as follows: In Section II, we derive the Lindblad equation for an open quantum system and for a theory which corrects the quantum mechanics. Then we describe some properties which are needed to get Born rule in the measurement.  In Section III, we present the Ramsey interferometer which is the base of atomic clocks in the observation and describes how corrections to quantum mechanics modify the excitation probability.
    In this review we shall assume $\hbar=1$.\\

    %%%%%%%%%%%%%%%%%%%%%%%%%%%%%%%%%%%%%%%%%%%%%%%%%%%
    \section{Lindblad equation as a quantum correction}
    %%%%%%%%%%%%%%%%%%%%%%%%%%%%%%%%%%%%%%%%%%%%%%%%%%%
    
    Probabilities may enter in quantum mechanics in two different ways: One is the source of mysterious spirit of quantum mechanics i.e. the probabilistic nature of state vectors; the other one is the classical probability related to the fact that we may not know the state vector of a given system. Consider a system in any one of a number of normalized states $|\Psi_i \rangle$ with classical probabilities $w_i$, which we call them weights. Also, suppose these state vectors are not necessarily orthogonal and that the probabilities are complete i.e $\sum_i w_i = 1$.
    In such cases, for quantum mechanical calculations, one finds the density matrix operator as a useful and applicable tool. It is defined as
    
    \begin{equation}\label{f1}
    \rho \equiv \sum_i w_i\;|\Psi_i \rangle \langle \Psi_i |.
    \end{equation}
    This operator has some definite properties: It is positive in the sense that all its eigenvalues are positive or equivalently $\langle \psi | \rho |\psi\rangle \geqslant 0$ for every vector $|\psi\rangle$. This operator is hermitian $\rho^\dagger = \rho$ and has a unit trace ${\rm Tr}\rho =1$.
    To find the expectation value of any observable represented by a Hermitian operator $A$,  we should first find the quantum mechanical expectation value of such operator $\langle \Psi_i | A |\Psi_i\rangle $ for each state vector and then calculate the mean value of these quantities with weights $w_i$. That is

    \begin{equation}\label{f2}
\langle A \rangle = \sum_i w_i\;\langle \Psi_i | A |\Psi_i \rangle = {\rm Tr} \Big( A\rho \Big) .
    \end{equation}
    Of course, at this stage, one has the right to ask "what happens after measurement?". According to the standard quantum mechanics, if the measuring is complete \footnote{We will describe the general version of measurement including complete and incomplete measurement.}, the initial density operator Eq. (\ref{f1}) collapses to a classical distribution of the eigenstates of the observable $A$, i.e. 
    
    \begin{equation}\label{f3}
    {\rho}_{initial} \equiv \sum_i w_i\;|\Psi_i \rangle \langle \Psi_i | \ \rightarrow \ \ {\rho}_{final} = \sum_m p_m\;|a_m \rangle \langle a_m |,
    \end{equation} 
    where $p_m = \langle a_m |     {\rho}_{initial} |a_m\rangle $ and $|a_m\rangle $ are eigenstates of $A$ and they make an orthonormal complete basis in the sense that $\sum_m |a_m\rangle \langle a_m | = I$ (this is why we call it a complete measurement). The condition Eq. (\ref{f3}) is called {\it Born Rule}.
    
    Based on the standard quantum mechanics, in a  system with a given Hamiltonian $H(t)$, each individual state vector $ \lvert \Psi(t) \rangle$ evolves according to the following first order differential equation

    \begin{equation}\label{f4}
    i \frac{d}{dt} \lvert \Psi(t) \rangle = H(t)\  \lvert \Psi(t) \rangle.
    \end{equation}
    This differential equation has a simple solution which may be written as:
    
    \begin{equation}\label{f5}
    \lvert \Psi(t) \rangle = U(t, t^{\prime}) \lvert \Psi(t^{\prime}) \rangle.
    \end{equation}
Here, $U(t, t^{\prime})$ is an operator that translates the initial state vector at time $t^{\prime}$ to any later time and it is clear that $ U(t^{\prime}, t^{\prime})=I$. If we put Eq. (\ref{f5}) into Eq. (\ref{f4}),  we shall find a differential equation for  $U(t, t^{\prime})$ in the form 

\begin{equation}\label{f6}
i \frac{d}{dt} U(t, t^{\prime}) = H(t)\  U(t, t^{\prime}).
\end{equation}

Using this equation, the initial condition $ U(t^{\prime}, t^{\prime})=I$, and the Hermitian condition of $H(t)$, one can easily show that $U(t, t^{\prime})$ is a unitary operator and so the state of the system evolves unitarily. With this description, it is easy to see that the density operator of a  system in Eq. (\ref{f1}) evolves as

 \begin{equation}\label{f7}
 \rho(t)= U(t, t^{\prime}) \rho(t^{\prime}) U^{\dagger}(t, t^{\prime}).
 \end{equation}
    Differentiating the above equation one obtains the dynamical equation for the density operator 
    
    \begin{equation}\label{f8}
    \dot{\rho} (t) = -i\left[ H(t), \rho(t) \right].
    \end{equation}
    The unitary evolution Eq. (\ref{f7}) does not lead $\rho_{initial}$ to $\rho_{final}$ in Eq. (\ref{f3}). To see this, let $\rho_{initial} = |\Psi \rangle \langle \Psi |$ be a pure state, then the unitary transformation Eq. (\ref{f7}) maps this pure state to another pure state, and not to the mixed state Eq. (\ref{f3}). In the standard quantum mechanics (with the original Copenhagen interpretation) people  accepted (as a principle) that the collapse dynamics departs from quantum mechanics. We shall return to this mysterious principle later on. 
    Are all systems supposed to have a unitary transformation in time? At least for some open systems, Eq. (\ref{f8}) is no longer valid. For example consider a system including two parts $\mathcal{S} + {\mathcal{S}}^{\prime}$. While the density matrix of the whole system evolves unitarily, in general, the density matrix of the subsystem $\mathcal{S}$ evolves in a non-unitary way(because of the influence of ${\mathcal{S}}^{\prime}$). Furthermore, sometimes we encounter a system influenced by environmental fluctuations or noises. In fact, the unitary evolution Eq. (\ref{f7}) is a special case of a general linear transformation which gives the components of the density matrix at a later time $t$ as a linear combination of the components of the density matrix at an earlier time $t^{\prime}$, with coefficients that are only functions of elapsed time $t - t^{\prime} $. That is

 \begin{equation}\label{f9}
 \rho_{ij} (t) = \sum_{i^{\prime} j^{\prime}} K_{i i^{\prime}, j j^{\prime}}(t- t^{\prime})\ \rho_{i^{\prime} j^{\prime}} (t^{\prime}).
 \end{equation}
Usually people call such an evolution as a kind of {\it Markovian evolutions}. Dependence on the elapsed time usually happens when the statistical properties of the system and its environment are independent of time. Such statistical properties in general lead us to a dynamical equation which is invariant with respect to time translation, and in Eq. (\ref{f9}) we can see this (It is invariant under a shift $t_0$ in $t$ and $t^{\prime}$). Here we have assumed that the dimension of the Hilbert space is finite and equal to $d$, but it can be extended even  to an infinite dimension.

To see an example where we face with Eq. (\ref{f9}), consider a system with a rapidly and randomly fluctuating Hamiltonian $H(t)$. If we look at the density matrix of such a system in the time scale for which fluctuations change, then the density matrix changes in time according to Eq. (\ref{f7}). But usually the density matrix changes very slowly in the characteristic time of fluctuations and an observer only distinguishes the average of the density matrix over fluctuations. In such cases we have

 \begin{equation}\label{f10}
 <\rho(t)>\  =\  < U(t, t^{\prime}) \rho(t^{\prime}) U^{\dagger}(t, t^{\prime}) >,
 \end{equation}
 and the kernel in Eq. (\ref{f9}) is given as
 \begin{equation}\label{f11}
 K_{i i^{\prime}, j j^{\prime}}(t- t^{\prime}) \equiv\  < U_{i i^{\prime}}(t, t^{\prime})\   U_{j j^{\prime}}^{\dagger}(t, t^{\prime}) >.
 \end{equation}
 Now, let us focus on Eq. (\ref{f9}). Because the transformation in Eq. (\ref{f9}) should transform a hermitian operator to another Hermitian operator, the kernel has to be Hermitian i.e

 \begin{equation}\label{f12}
 K_{i i^{\prime}, j j^{\prime}}(t- t^{\prime}) = K^{*}_{ j j^{\prime}, i i^{\prime}}(t- t^{\prime}).
 \end{equation}
 
 Also, this kernel should leave the trace of the density matrix invariant. Under this condition, we find the following relation 
   
 \begin{equation}\label{f13}
\sum_{i} K_{i i^{\prime}, i j^{\prime}}(t- t^{\prime}) = \delta_{i^{\prime} j^{\prime}}.
 \end{equation}
  From here we shall replace $t-t^\prime$ with $\tau$. The condition Eq. (\ref{f12}) tells us that we can diagonalize $K_{i i^{\prime}, i j^{\prime}}$ and decompose it to its eigenvectors

 \begin{equation}\label{f14}
 K_{i i^{\prime}, j j^{\prime}}(\tau) = \sum_{N=1}^{d^2} \alpha_{N}(\tau)\ u^{N}_{i i^{\prime}}(\tau)\ {u^{N}_{j j^{\prime}}}^{*}(\tau),
 \end{equation}
where $\alpha_{N}(\tau)$ are real eigenvalues of $ K_{i i^{\prime}, j j^{\prime}}(\tau)$, with the corresponding eigenvectors (which are matrices) $u^{N}_{i i^{\prime}}(\tau)$ i.e.

\begin{equation}\label{f15}
\sum_{j j^{\prime}} K_{i i^{\prime}, j j^{\prime}}(\tau) u^{N}_{j j^{\prime}}(\tau) = \alpha_{N}(\tau) u^{N}_{i i^{\prime}}(\tau).
\end{equation}

These eigenvectors should be orthonormal and so 
    
\begin{equation}\label{f16}
\sum_{i i^{\prime}}  u^{N}_{i i^{\prime}}(\tau)\ {u^{M}_{i i^{\prime}}}^{*}(\tau) = \delta_{N M}.
\end{equation}
If we use the decomposed form of kernel in Eq. (\ref{f14}) and the trace condition Eq. (\ref{f13}), we find

\begin{equation}\label{f17}
\sum_{N} \alpha_{N}(\tau){u^{N}}^{\dagger}(\tau)\  u^{N}(\tau) = I.
\end{equation}

One should note that the kernel is like a $d^2 \times d^2$ Hermitian matrix. Hence, the number of its independent eigenvectors is $d^2$. By the using  Eq. (\ref{f14}) , one may rewrite Eq. (\ref{f9}) in the following form

 \begin{equation}\label{f18}
 \rho_{ij} (t) = \sum_{i^{\prime} j^{\prime}} \sum_{N} \alpha_{N}( t- t^{\prime})\ u^{N}_{i i^{\prime}}( t- t^{\prime})\ {u^{N}_{j j^{\prime}}}^{*} (t- t^{\prime})\ \rho_{i^{\prime} j^{\prime}} (t^{\prime})
 \end{equation}
 or in a more abstract form
 
 \begin{equation}\label{f19}
 \rho (t) =\sum_{N} \alpha_{N}( t- t^{\prime})\ u^{N}( t- t^{\prime})\ \rho (t^{\prime}) {u^{N}}^{\dagger}(t- t^{\prime}).
 \end{equation}
 
At this stage, we shall try to find a differential equation for the density matrix of an open system which its dynamics admits Eq. (\ref{f9}). For this purpose, we shall use the first order perturbation theory (See appendix A).  To work this out, we need to look at the neighborhood of $t^\prime$ and investigate $K$ and its eigenvectors and eigenvalues in this neighborhood. Eq. (\ref{f9}) tells us that when $t^\prime = t$, we have 
       
 \begin{equation}\label{f20}
 K_{i i^{\prime}, j j^{\prime}}(0) =\delta_{i i^{\prime}} \delta_{j j^{\prime}} .
 \end{equation}
 This operator admits an eigenvector with eigenvalue equal to $d$, which we label them with $N=1$:
 
 \begin{equation}\label{f21}
  u^{1}_{i i^{\prime}}(0) = \frac{1}{\sqrt{d}}\delta_{i i^{\prime}} ,\ \ \ \ \ \ \alpha_{1}(0)= d,
 \end{equation}
and the rest of eigenvectors are the degenerate with the eigenvalue zero which we label them with $N=a$. To be eigenvectors they should satisfy  

 \begin{equation}\label{f22}
 Tr[ u^{a}(0)]=0 ,\ \ \ \ \ \ \alpha_{a}(0)= 0, \ \ \ \ \ a= 2 \ldots d^2.
 \end{equation}
 
 That is, these eigenvectors must be traceless matrices and the number of them is $d^2 -1$. Here one should be careful when one uses the first order perturbation theory. As is shown in Appendix A, when an operator in the zeroth order has some degenerate eigenvectors with an eigenvalue, in order for the eigenvectors at the zeroth order be connected smoothly with the first order corrections, these degenerate eigenvectors must be chosen such that the operator at its first order correction has a diagonal form with respect to these eigenvectors. Therefore, for small $\tau \in R^+$, in order for the eigenvectors $u^a(0)$ to have a smooth connection with eigenvectors $u^a(\tau)$ of $K(\tau)$,  they are chosen not only to be  traceless but  they also diagonalize the Kernel $K(\tau)$ in its first order correction. That is 
     
 \begin{equation}\label{f23}
\sum_{i i^{\prime} j j^{\prime}}\ u^{a}_{j j^{\prime}}(0)\ {u^{b}_{i i^{\prime}}}^{*}(0)\ \left[\frac{d  K_{i i^{\prime}, j j^{\prime}}(\tau)}{d\tau}\right]_{\tau=0} \ =\ \delta_{ab}\ \left[\frac{d\alpha_{a}(\tau)}{d\tau}\right]_{\tau=0}.
 \end{equation}
 
 Now, we can come back to Eq. (\ref{f9}). Suppose the elapsed time is very small i.e. $t-t^\prime =\epsilon$. Using the Tylor expansion for the right hand side (RHS) and the left hand side (LHS), we shall find the following differential equation 
 
 \begin{equation}\label{f24}
\dot{\rho} (t) =\sum_{a} \eta_{a}\ u^{a}(0)\ \rho (t) {u^{a}}^{\dagger}(0)\  +\ A\rho(t) + \rho(t) A^{\dagger},
 \end{equation}
 where

 \begin{equation}\label{f25}
  A \equiv \sqrt{d} \left[\frac{d u^{1}(\tau)}{d\tau}\right]_{\tau=0} + I\ \left[\frac{1}{2d}\frac{d\alpha_{1}(\tau)}{d\tau}\right] _{\tau=0} 
 \end{equation}
 and $\eta_{a} \equiv \left[\frac{d\alpha_{a}(\tau)}{d\tau}\right]_{\tau=0}$. It would be possible to find a more convenient form for Eq. (\ref{f24}). To do this, it is enough to take the derivative of Eq. (\ref{f17}) at $\tau =0$. This will give: 
 \begin{equation}\label{f26}
 \sum_{a} \eta_{a}\ u^{a}(0) {u^{a}}^{\dagger}(0) + A +  A^{\dagger} = 0,
 \end{equation}
 which determines the hermitian part of $A$. One can always decompose a matrix into the sum of its hermitian and anti-hermitian parts. Define the anti-Hermitian part of $A$ as $- i{\mathcal{H}}^{\prime}$, we have:
 
\begin{equation}\label{f27}
A = -\frac{1}{2}\sum_{a} \eta_{a}\ u^{a}(0) {u^{a}}^{\dagger}(0) - i{\mathcal{H}}^{\prime}.
\end{equation}
Thus, Eq. (\ref{f24}) can be written as 

\begin{equation}\label{f28}
\dot{\rho} (t) = -i\left[ {\mathcal{H}}^{\prime}, \rho(t) \right] + \sum_{a} \eta_{a}\ \left[ u^{a}(0)\ \rho (t) {u^{a}}^{\dagger}(0)\  - \frac{1}{2}\rho(t) {u^{a}}^{\dagger}(0) u^{a}(0)  - \frac{1}{2} {u^{a}}^{\dagger}(0) u^{a}(0) \rho(t) \right].
\end{equation}
Here, we have a differential equation with the matrices $u^a(0)$ which should be traceless. One can make more progress and drop out this constraint, by the following redefinition
 
 \begin{equation}\label{f29}
u^{a}(0) \equiv c_{a}^{*} I +  L^{\prime}_a,
 \end{equation}
 where $L^{\prime}_a$ are  arbitrary matrices and $c_a$ is a complex number with value $-{\rm Tr (\frac{1}{d} L^{\prime}_a)}$. Substituting Eq. (\ref{f29}) in Eq. (\ref{f24}), we get

 \begin{equation}\label{f30}
 \dot{\rho} (t) = -i\left[ \mathcal{H}, \rho(t) \right] + \sum_{a} \eta_{a}\ \left[ L^{\prime}_a\ \rho (t) {L^{\prime}_a}^{\dagger}\  - \frac{1}{2}\rho(t) {L^{\prime}_a}^{\dagger} L^{\prime}_a  - \frac{1}{2} {L^{\prime}_a}^{\dagger} L^{\prime}_a \rho(t) \right],
 \end{equation}
where
 \begin{equation}\label{f31}
\mathcal{H} \equiv  {\mathcal{H}}^{\prime} + \frac{i}{2} \sum_{a} \eta_{a}\left[ \ c_{a}^{*} {u^{a}}^{\dagger}(0) - c_{a} u^{a}(0)  \right].
 \end{equation}
 Note that $\mathcal{H}$ is a Hermitian operator. Now, we have a differential equation without any constraint on $L^{\prime}_a$. 
 
 Up to this level, we have used two essential condition for the density matrix transformation Eq. (\ref{f9}): If $\rho(t^\prime)$ on the RHS is a matrix with a unite trace, then  $\rho(t)$ on the LHS should also have unit trace. If $\rho(t^\prime)$ is a Hermitian matrix, then  $\rho(t)$ should inherit this property. There still exists another property that we have not used i.e. the positivity of $\rho$. Under what condition on the kernel $K(\tau)$, the positivity of $\rho(t^\prime)$ grantees the positivity of $\rho(t)$ in Eq. (\ref{f9})? To answer this question, let us take a closer look at Eq. (\ref{f19}). Sandwich this equation from both sides with an arbitrary vector $|v\rangle$:
 \begin{equation*}
\langle v|\rho (t) |v\rangle=\sum_{N} \alpha_{N}( t- t^{\prime})\ \langle v|u^{N}( t- t^{\prime})\ \rho (t^{\prime}) {u^{N}}^{\dagger}(t- t^{\prime})|v\rangle = \sum_{N} \alpha_{N}( t- t^{\prime})\ \langle w^a(t- t^{\prime})| \rho (t^{\prime}) |w^a(t- t^{\prime})\rangle,
 \end{equation*}
where $|w^a(t- t^{\prime})\rangle\equiv {u^{N}}^{\dagger}(t- t^{\prime})|v\rangle$.  Apparently when $\alpha_{N}$ are non-negative numbers, $\rho(t)$ will be positive, but the inverse is not necessary true. It is plausible for $K(\tau)$ to have negative eigenvalues  while it preserves the positivity condition for the density matrix \footnote{Some authors use the transpose operation $K_{ij,mn}= \delta_{jm}\delta_{in}$ as an example of such cases. Because for the transpose operator we have $K^2=I$ this operator has eigenvalues $\pm 1$ and maps a positive matrix to a positive matrix. Unfortunately such an example is irrelevant to Eq. (\ref{f9}) because it obviously violates Eq. (\ref{f20}). }. The eigenvalue $\alpha_{1}(\tau)$ has the value $d$ at $\tau=0$ and even if $\dot{\alpha}_{1}(0)$ is a negative number, there is always a neighborhood of $\tau=0$ (for $\tau \in R^+$) in which  $\alpha_{1}(\tau)$ is positive. While $\alpha_{a}(\tau)$ at are equal to zero  $\tau=0$, there non-negativity  in a neighborhood of $\tau=0$ would be guaranteed if $\eta_{a} =\dot{\alpha}_{a}(0)$ have some non-negative values. If this condition is satisfied, then we can rewrite Eq. (\ref{f30}) as

 \begin{equation}\label{f32}
 \frac{d\rho(t)}{dt}=-i[{\cal H},\rho(t) ]+\sum_a \left[L_a\rho(t)L_a^\dagger-\frac{1}{2}L_a^\dagger L_a \rho(t)
 -\frac{1}{2}\rho(t) L_a^\dagger L_a\right]\;,
 \end{equation}
 where $L_a \equiv \sqrt{\eta_{a}} L^{\prime}_a$. There is a mathematical theorem that explains under what condition all $\alpha_{N}$ in the transformation Eq. (\ref{f19}) are non-negative. This theorem states that:\\
 
 \textit{Under the transformation Eq. (\ref{f19}) of the positive matrices, all $\alpha_{N}$ are non-negative if and only if the transformation Eq. (\ref{f19}) is a completely positive transformation}.\\

  Here we assume that the reader is familiar with the complete positivity concept. If it is not the case, we have provided a short appendix (Appendix C) at the end of this review. We encourage the reader to study Appendix C and then return to the main line.
 
  Equation (\ref{f32}) is usually called {\it Kossakowski-Lindblad equation}, which we briefly call it Lindblad equation in this review.

 The Lindblad equation can also be derived when one tries to find the reduced dynamics of systems \cite{spoh} which are in a weak interaction with their environment. Benatti, Floreanini, and Romano (BFR) wrote their paper \cite{ben}, when there was a debate about whether the complete positivity condition is physically necessary or not. In quantum communication theory, where people deal with local operations on quantum states, it should be necessary to consider the evolution of states as a  completely positive map. To be more concrete, consider a bipartite system which is in the Bell state $|\psi\rangle= \frac{1}{\sqrt{2}}\Big(|+\rangle_A \otimes |-\rangle_B + |-\rangle_A \otimes |+\rangle_B  \Big)$, where $|\pm\rangle$ are orthonormal vectors. Consider the first sector of this state to be available for Alice and the second sector belongs to Bob. The whole system $AB$ is described by the density matrix $\rho_{AB} = |\psi\rangle \langle\psi |$. If Alice makes an experiment on her sector while Bob keeps his own sector intact, Alice's action on $\rho_{AB}$ is shown by operator $\mathcal{E} \otimes I$, where $\mathcal{E}$ is Alice's action on her sector. This operator must map $\rho_{AB}$ to another density matrix and because we can consider any dimension for Bob's Hilbert space, $\mathcal{E}$ should be a completely positive map. Of course, such argument would be useful(and valid) when there are physical Hilbert spaces of any dimension which are invariant under the time evolution. But in our world, there are no such Hilbert spaces, except the vacuum that constructs only a one-dimensional Hilbert space. \\

   %%%%%%%%%%%%%%%%%%%%%%%%%%%%%%%%%%%
   \subsection{Complete positivity}
   %%%%%%%%%%%%%%%%%%%%%%%%%%%%%%%%%%%%%%%%%%%
   
   In 2002, BFR found a condition that is physically necessary and grantees the complete positivity of Eq. (\ref{f9}). To understand their argument, first let us re-express Eq. (\ref{f9}) in a more abstract way. Suppose $\gamma_t$ be a continuous linear map from space of density matrices ( of dimension $d$) $M_d(C)$ to itself
\begin{equation*}
\gamma\ :\ R^+ \times M_d(C) \longrightarrow\  M_d(C)
\end{equation*} 
We can consider the kernel $K(t)$ in Eq. (\ref{f9}) as a matrix representation of this map and rewrite Eq. (\ref{f9}) in the new form $\rho(t) = \gamma_t (\rho)$. Here for simplicity we take $t^\prime =0$, and note that $K(t)$ has this property that while it transforms a density matrix at time $t_1$ to another one at a later time $t_2 = t_1 + \tau_1$\ ($\tau_1 \geqslant 0$), it is only function of the elapsed time $\tau_1$. Therefore, if we like to have the density matrix at another time $t_3 = t_2 + \tau_2$\ ($\tau_2 \geqslant 0$), it is enough to act $K(\tau_1 + \tau_2)$ on $\rho(t_1)$. Thus if we want $\gamma_t$ to be equivalent with the positive trace-preserving kernel $K(t)$, it should satisfy the following conditions:

 \begin{eqnarray}
 \label{f33}
 \gamma_t \circ \gamma_s &=& \gamma_{t + s} = \gamma_s \circ \gamma_t\ ,
 \quad\forall s,t \geqslant 0\ ; \\
 \label{f34}
 {\rm Tr}\ \gamma_t[\rho] &=& {\rm Tr}\ \rho,\ \; \gamma_t[\rho]^{\dagger} =
 \gamma_t[\rho]\ ; \\
 \label{f35}
 \lim_{t\rightarrow 0^+}\gamma_t[\rho] &=& \rho\ ,
 \end{eqnarray}
  People call such maps $\{\gamma_t\}$ for $t\geqslant 0$ as semigroup of positive linear maps(Because they form a group without the inverse elements). There is a famous theorem \cite{gori} which states that any semigroup $\{\gamma_t\}_{t\geqslant 0}$ satisfying conditions Eq. (\ref{f33}), Eq. (\ref{f34}) and Eq. (\ref{f35}), should be generated by 

  \begin{equation}\label{f36}
  \partial_t \gamma_t (\rho)=\partial_t \rho(t) = -i \left[H, \rho(t)\right] 
  + \sum_{m,n
      = 1}^{d^2 -1} c_{mn}\left[F_m \rho(t) F_n^{\dagger}
  - \frac{1}{2} \{F_n^{\dagger} F_m, \rho(t)\}\right]\ ,
  \end{equation}
  where $H=H^\dagger$,\ ${\rm Tr}\Big(F_m F_n^\dagger\Big) = \delta_{mn}$,\ ${\rm Tr}F_m = 0$, and $c_{ab}$ are components of a $(d^2-1)\times (d^2-1)$ Hermitian matrix, depending on the choices of $F_m$. This equation is equivalent with what we have in Eq. (\ref{f28}). To see this, we note that because $C=[c_{ab}]$ is a Hermitian matrix, it can be written as $c_{mn} =
 \sum_{r = 1}^{d^2 - 1}  \eta_a R^*_{am} R_{an}$. If we put it into Eq. (\ref{f36}) and define $L_a^{\prime} = \sum_{m = 1}^{d^2 -1} R_{am}^* F_m$, we shall reach Eq. (\ref{f28}). For a completely positive semigroup $\{\gamma_t\}_{t\geqslant 0}$, it is clear that $C$ must be a positive definite matrix (Because its eigenvalues $\eta_a$ are non-negative)  \footnote{And vice versa, if $C$ is a positive definite matrix then the semigroup $\{\gamma_t\}_{t\geqslant 0}$ should be  completely positive .} and can be written in form of $C=A^\dagger A$. Therefore, with the redefinition $ L_a = \sum_{m = 1}^{d^2 -
     1} A_{am}^* F_m$, we find Lindblad equation. But how can we argue that {\it for physical systems $C$ is positive definite}? 
 To answer this question, we turn our attention to two points : first, if a system $\mathcal{S}$ would be physically realizable, then the combined system $\mathcal{S}\oplus \mathcal{S}$ consisting of two isolated  copies  of $\mathcal{S}$ should also be so(by isolated we mean there are no interactions between two copies, but they separately interact with the same environment). Second, if the dynamics of $\mathcal{S}$ is governed by $\{\gamma_t\}_{t\geqslant 0}$(or equivalently by $K(t)$), then the dynamics of the whole system  $\mathcal{S}\oplus \mathcal{S}$ (the isolated copies) should be described by $\{\gamma_t \otimes \gamma_t \}_{t \geqslant 0}$ (or equivalently by $K(t)\otimes K(t)$). In other words,  if $\rho$ is an arbitrary density matrix of the combined system over the whole Hilbert space $H\otimes H$, then for every $|\phi\rangle \in H\otimes H$, we should have
 \begin{equation}\label{f37}
  \langle \phi|\{\gamma_t \otimes \gamma_t \} (\rho) |\phi\rangle \geqslant 0.
   \end{equation}
   We are careful about the fact that for the positive maps $\gamma_t$ this condition is non-trivial when $\rho$ is an entangled density matrix. If $\rho = \rho_1 \otimes \rho_2$, then $\{\gamma_t \otimes \gamma_t \}_{t \geqslant 0}$ maps such state to another positive matrix. With this consideration, let define the function $g(t)$ as following
  
  \begin{equation}\label{f38}
 g(t)\equiv  \langle \phi \vert (\gamma_t \otimes \gamma_t) 
 [\vert \psi \rangle
 \langle \psi \vert ] \vert \phi \rangle  \geqslant 0, 
 \end{equation}
  where we have set $\rho \rightarrow |\psi\rangle \langle \psi |$ in Eq. (\ref{f37}) and have chosen $|\phi\rangle$ and $|psi\rangle$ orthogonal. These vectors have matrix representations $\Psi$ and $\Phi$ in basis $\{|i\rangle \otimes |j\rangle  \}_{i,j=1}^{d^2-1}$ :

  \begin{equation}\label{f39}
  \vert \phi \rangle = \sum_{j,k = 1}^{d}{[\Phi]}_{jk}\  \vert j
  \rangle \otimes \vert k \rangle;\ \ \ \  
  \vert \psi \rangle = \sum_{j,k =
      1}^{d} {[\Psi]}_{jk}\  \vert j \rangle \otimes \vert k \rangle.
  \end{equation}
  
For the small value of $t$ we have

\begin{equation}\label{f40}
g(t)\approx g(0) + t\Big[\frac{dg(t)}{dt}\Big]_{t=0}  = t\Big[\frac{dg(t)}{dt}\Big]_{t=0} ,\ \ \ \ \ g(t) \geqslant 0 \ \rightarrow \Big[\frac{dg(t)}{dt}\Big]_{t=0} \geqslant 0.
\end{equation}

Using Eq. (\real{f36}), and after some simple ordinary calculations, one will find that  
 
\begin{equation}\label{f41}
\Big[\frac{dg(t)}{dt}\Big]_{t=0} = 
\sum_{m,n = 1}^{d^2 - 1} c_{mn} \left[{\rm Tr}\Big(\Psi
\Phi^{\dagger}F_m\Big)\ {\rm Tr}\Big(\Phi \Psi^{\dagger}F_n^{\dagger}\Big) +  
{\rm Tr}\Big( (\Phi^{\dagger}\Psi)^T F_m\Big)\ {\rm Tr}\Big( (\Psi^{\dagger} \Phi)^T
F_n^{\dagger}\Big) \right]\ .
\end{equation}
 Suppose $\{ w_m\}_{m=1}^{d^2-1}$ are some arbitrary c-numbers and let us define the traceless matrix $W\equiv \frac{1}{2}\sum w^{*} F_m$. A lemma in linear algebra states that every matrix $W$ is similar to its transposed $W^{T}$. Therefore, there always exists a non-singular matrix $U$ such that $W^T = U^{-1} W U$. If we set $\Phi=U$ and $\Psi^\dagger = U^{-1} W$, one easily sees that $\Phi \Psi^\dagger = W$ and $W^T = \Psi^\dagger \Phi$. Substituting these relations in Eq. (\ref{f41}), we shall find 
 \begin{equation*}
 \sum_{m,n=1}^{d^2-1} c_{mn} w^{*}_{m} w_n \geqslant 0.
 \end{equation*}
 Thus, $C$ is a positive definite matrix and $\{ \gamma_t\}_{t \geqslant 0}$ must be completely positive.  
 
Before we go further and find the solution of the Lindblad equation, we like to mention an important point about Eq. (\ref{f9}). As Gisin showed \cite{gisi} in his study of the entangled states, in order to avoid instantaneous communication at a distance (inconsistency with relativity) it is necessary that the density matrix at a given time $t$ to depend on the density matrix at any earlier time $t^\prime \leq t$, but not on the state vector at $t$, thus, in general Eq. (\ref{f9}) has no inconsistencies with relativity. An open system $\mathcal{S}$ in the standard quantum mechanics may interact with its environment $E$ which includes some parameters fluctuating randomly and more rapidly than the rate at which the density matrix of $\mathcal{S}$ changes. Averaging over these parameters can lead us to a linear but non-unitary evolution i.e. Eq. (\ref{f32}). In finding such an evolutionary equation, we didn't need to know the details of these environmental fluctuations; We only considered the physical conditions for this evolution: This should be a completely positive trace-preserving Hermitian map. It is interesting that the Lindblad equation not only appears in open quantum systems of standard quantum mechanics but also can describe the dynamics of a wide range of collapse models like the CSL (Continuous Spontaneous Localization) model of the modified quantum theory. \\

%%%%%%%%%%%%%%%%%%%%%%%%%%%%%%%%%%%%%%%%%%
\subsection{Lindblad equation and Born rule}
%%%%%%%%%%%%%%%%%%%%%%%%%%%%%%%%%

At this point we shall follow Weinberg \cite{Weinberg:2016axv} and investigate the solution of Eq. (\ref{f32}) to answer the question that : \textit{Does the Lindblad equation admit  Born rule?} The answer is yes, but as we shall see, to have  Born rule Eq. (\ref{f3}) there should be a physical condition on  $L_a$ and $\mathcal{H}$ in Eq. (\ref{f32}). What we want is the solution of Eq. (\ref{f32}) at a late time approaches to a time-independent linear combination of the specific projection operators $P_m = |a_m \rangle \langle a_m |$ of an operator $A$ for any initial condition $\rho_{initial}$. The coefficients in this linear combination should be given by the probabilities $p_m = \langle a_m| \rho_{initial} | a_m \rangle$. Eq. (\ref{f32}) is a linear equation with time-independent coefficients and can be put in the following form

 \begin{equation}\label{f42}
\frac{d\rho(t)}{dt} = {\cal L}\rho(t),
 \end{equation}
 where $\cal{L}$ is a $(d^2 \times d^2)$ time-independent (super) matrix, acting on the space of $d\times d$ matrices and is defined as
 
  \begin{equation}\label{f43}
  {\cal L}\rho\equiv -i[{\cal H},\rho]+\sum_a \left[L_a\,\rho\,L_a^\dagger-\frac{1}{2}L_a^\dagger L_a\, \rho
 -\frac{1}{2}\rho\, L_a^\dagger L_a\right]\;.
 \end{equation}
  For the moment, let us suppose that $\cal{L}$ is a diagonalizable matrix with eigenvalues $-\mu_n$ and corresponding right eigenvectors $\rho_n$ 
  
 \begin{equation}\label{f44}
 {\cal L}\rho_n = -\mu_n\  \rho_n\;.
 \end{equation}
 Then, the generic solution of Eq. (\ref{f42}) would be
 
 \begin{equation}\label{f45}
 \rho(t)=\sum_n \rho_n\ e^{-\mu_n t}\;.
 \end{equation}
 
 The number of  linearly independent eigenvectors $\rho_n$ is $d^2$ (  for non-diagonalizable $\cal{L}$ this is less than $d^2$). Because $\{ \rho_n\}_{n=1}^{d^2}$ is a complete basis (they are orthogonal but not necessary normalized), the normalization of each $\rho_n$ depends on the initial condition $\rho_{initial}$ \footnote{For the usual inner product  on the space of $d\times d$ matrices these normalizations are given by ${\| \rho_n \|}^2 = {\rm Tr}(\rho^{\dagger}_n \rho_{initial})$}. We note that because $\cal{L}$ is not in general Hermitian, its eigenvalues are complex numbers. Although the sum in Eq. (\ref{f45}) should be Hermitian, positive and trace $1$, but these conditions are not necessary for each $\rho_n$. From definition Eq. (\ref{f43}), it is clear that $\cal{L}$ has a trivial right eigenvector proportional to $I_{d^2 \times d^2}$, with eigenvalue zero. We can even go further and argue that $I_{d^2 \times d^2}$ is also a left eigenvector of any trace-preserving operator $\cal{L}$, satisfying Eq. (\ref{f42}). To see this, we write Eq. (\ref{f42}) as $\dot{\rho}_{ij}(t) =\sum_{mn} {\cal{L}}_{ij;mn}\rho_{mn}(t)$. The trace preserving condition tells us 
 \begin{equation*}
 \sum_{ij} \delta_{ij} \dot{\rho}_{ij}(t) = 0= \sum_{mn}\Big(\sum_{ij} \delta_{ij} {\cal{L}}_{ij;mn} \Big) \rho_{mn}(t)\   \Longrightarrow \ \sum_{ij} \delta_{ij} {\cal{L}}_{ij;mn} = 0.
 \end{equation*}
  This shows that (independent of definition Eq. (\ref{f43})) $\cal{L}$ is a singular operator (its determinant vanishes) and therefore has at least one right eigenvector with zero eigenvalue(not necessarily $I_{d^2 \times d^2}$ because in deriving $det \mathcal{L} = 0$, we did not use definition Eq. (\ref{f43})).
  The operator $\mathcal{L}$ may have some eigenvalues with positive-definite real part (Re$ (\mu_n) < 0$). It is important to note that the initial condition for the density matrix rules out such eigenvalues with their corresponding eigenvectors to have any contribution in the solution Eq. (\ref{f45}) . In other words, such eigenvectors live in a space which is normal to space of density matrices\footnote{This normal space is defined as the space of all matrices $A$ satisfying ${\rm Tr}(A^\dagger \rho)=0$ for all density matrices. It is easy to show that it is a vector space.}. Suppose such eigenvectors contribute to Eq. (\ref{f45}) and let us define the sum of such terms as $R(t)$. Then $\rho(t) \rightarrow R(t)$ for large $t$. But we know that ${\rm Tr}\rho(t) =1$ at all times and this will be possible when we have ${\rm Tr}R(t) =0$. On the other, hand the density matrix should remain positive and hermitian at all times, so we expect $R(t)$ to be a positive and hermitian matrix. But the only traceless and positive hermitian matrix is zero matrix and therefore positive-definite eigenvalues have no contributions in Eq. (\ref{f45}). In fact, the terms with the eigenvalues of the negative-definite real part would be suppressed exponentially and only the sum of terms with Re$(\mu_n)=0$  dominates $\rho(t)$ at the late time in Eq. (\ref{f45}).
   
  By multiplying LHS and RHS of Eq. (\ref{f44}) with $\rho_n$ and then taking the trace, we have

 \begin{equation}\label{f46}
{\rm Tr}\Big[\rho_n^\dagger {\cal L}\rho_n\Big] =-\mu_n {\rm Tr}\Big[\rho_n^\dagger \rho_n\Big] \;.
 \end{equation}
 If we use the explicit form of $\cal{L}$ in Eq. (\ref{f43}), after a straightforward calculation, we can separate the real and imaginary part of Eq. (\ref{f46}) as

 \begin{equation}\label{f47}
{\rm Tr}\Big[\rho_n^\dagger \rho_n\Big]{\rm Re}(\mu_n) =\frac{1}{2}{\rm Tr}\left(\sum_a [\rho_n\,,\,L_a^\dagger]^\dagger[\rho_n\,,\,L_a^\dagger]\right)
+ \frac{1}{2}{\rm Tr}\left( \sum_a\Big[L_a^\dagger L_a-L_a L_a^\dagger\Big] \rho_n \rho_n^\dagger\right) \;
 \end{equation}
 
 and

 \begin{equation}\label{f48}
{\rm Tr}\Big[\rho_n^\dagger \rho_n\Big]{\rm Im}(\mu_n)= - {\rm Im}\left( {\rm Tr}\sum_a L_a \rho_n^\dagger[\rho_n,L_a^\dagger ]\right) + {\rm Tr}\Big(\rho_n^\dagger [{\cal H},\rho_n]\Big)
 \;.
 \end{equation}
At this step we would like to invoke an assumption that limits the Lindblad operators $L_a$. Suppose $L_a$ satisfies the following condition
\begin{equation}\label{f49}
\sum_a \Big(L_a^\dagger L_a - L_a L_a^\dagger \Big) = 0.
\end{equation}
We shall return to the implication of this condition soon, but for  moment we just assume it. This condition leads Eq. (\ref{f47}) to 

 \begin{equation}\label{f50}
 {\rm Tr}\Big[\rho_n^\dagger \rho_n\Big]{\rm Re}(\mu_n) = \frac{1}{2}{\rm Tr}\left(\sum_a [\rho_n\,,\,L_a^\dagger]^\dagger[\rho_n\,,\,L_a^\dagger]\right)\;.
 \end{equation}
 We see that under assumption Eq. (\ref{f49}) the real part of all  $\mu_n$ are non-negative. As discussed before, $\rho(t)$ at late time should be dominated by the linear combination of those eigenvectors for which Re$(\mu_n)=0$. Because we are interested in late time behavior of the density matrix, we shall only focus on eigenvectors with purely imaginary eigenvalues (including zero). By looking at Eq. (\ref{f50}) we see that each $\rho_n$ with a purely imaginary eigenvalue $\mu_n$ commutes with all $L_a^\dagger$. One can also easily argue that if $\rho_n$ is a right eigenvector of the operator $\cal{L}$ defined in Eq. (\ref{f43}), with a purely imaginary eigenvalue $\mu_n$, then $\rho_n^\dagger$ would also be a right eigenvector of $\cal{L}$ with purely imaginary eigenvalue $-\mu_n$. Therefore  $\rho_n^\dagger$ also commutes with all $L_a$. We see that for such eigenvectors Eq.(\ref{f48}) reduces to
 
 \begin{equation}\label{f51}
 {\rm Tr}\Big[\rho_n^\dagger \rho_n\Big]{\rm Im}(\mu_n)=  {\rm Tr}\Big(\rho_n^\dagger [{\cal H},\rho_n]\Big)
 \;.
 \end{equation}
 Moreover,  it is rather trivial to check from Eq. (\ref{f43}), these eigenvectors satisfy

 \begin{equation}\label{f52}
\mu_n \rho_n = i[{\cal H},\rho_n]\;.
 \end{equation}
 The inverse is also true. If a vector satisfies Eq. (\ref{f52}) and it commutes with all $L_a$, it is a right eigenvector of $\cal{L}$, with a purely imaginary  eigenvalue. Thus, we have the following lemma \\
  
  \textit{\textbf{Lemma: }Under condition Eq. (\ref{f49}) a vector $\rho_n$ is a right eigenvector of $\mathcal{L}$ with a purely imaginary eigenvalue $\mu_n$ if and only if it commutes with all $L_a$
and is an eigenvector of the adjoint representation of $\mathcal{H}$}.
\footnote{The adjoint representation $Ad_{A}$ of an operator $A$ (in a Lie vector space $\mathcal{V}$) is defined $Ad_{A}(V) \equiv [A, V]$ when acts on every arbitrary operator $V \in \mathcal{V}$.}\\

It is interesting that the space of such eigenvectors is closed under the Lie bracket $[,]$ and so defines a Lie algebra.

In general, $\cal{L}$ is not a diagonalizable matrix (See Appendix B) and so its eigenvectors cannot span the whole space on which $\cal{L}$ acts. In such a case, to solve Eq. (\ref{f42}) the non-trivial generalized eigenvectors of $\cal{L}$ are also needed. For  non-diagonalizable $\cal{L}$, each $\rho_n$ in the solution Eq. (\ref{f45}) turns out to be a polynomial in $t$ of the order $k$, if the corresponding eigenvalue has $k$ non-trivial generalized eigenvectors. We note that those purely imaginary eigenvalues having non-trivial generalized eigenvectors, do not contribute to Eq. (\ref{f45}) or if they contribute, the initial condition for the density matrix does not let them be accompanied with time-dependent coefficients. In fact, the non-trivial generalized eigenvectors of purely imaginary eigenvalues live in the normal space of density matrices. The argument would be similar to what we presented for eigenvalues with a positive-definite real part.

Now, let us come back to the measurement problem (the Born rule). We suppose that in a measuring process of an observable $A$, the macroscopic measuring apparatus interacts with the system under study in such a way that the Lindblad equation describes the evolution of the system. To drive Born rule,  we first need to find $\mathcal{H}$ and $L_a$ for which  $\rho(t)$ at late time, approaches a linear combination of the projection operator $P_m=|a_m\rangle\langle a_m|$,  of a specific observable $A$ for all initial conditions $\rho_{initial}$. As discussed before, for the solution Eq. (\ref{f45}), the summation will approach a linear combination of the eigenvectors with purely imaginary eigenvalues at late time. Let us consider this asymptotic limit as $\rho_f(t) \equiv \lim_{t \rightarrow \infty} \rho(t)$. To obtain the Born rule, this asymptotic limit should coincide with Eq. (\ref{f3}) for all $\rho_{initial}$. In other words
\begin{equation}\label{fo1}
\rho_f(t) = \sum_\alpha  p_\alpha P_\alpha,
\end{equation}
where $P_\alpha =|a_\alpha \rangle \langle a_\alpha | $ and $p_\alpha= \langle a_\alpha | \rho_{initial}|a_\alpha \rangle$. 
  Because all $L_a$ commute with the eigenvectors with the corresponding purely imaginary eigenvalues, $\rho_f(t)$ should also commute with all $L_a$. This means 
  
  \begin{equation*}
\Big[L_a, \sum_\alpha  p_\alpha P_\alpha \Big]=0.
  \end{equation*}
  Now, if we choose $\rho_{initial}= P_\beta$ for every $\beta$, we shall have
  
  \begin{equation}\label{f53}
  [L_a,P_\beta]=0\;. 
  \end{equation}
  
  This means that\footnote{To see this just note that for every $L_a$ satisfying Eq. (\ref{f53}), we have
$$
 L_a|a_\alpha\rangle=L_a P_\alpha|a_\alpha\rangle=P_\alpha L_a|a_\alpha\rangle=|a_\alpha\rangle \langle a_\alpha|L_a|a_\alpha\rangle\;.
$$
In other words, $L_a$ is diagonalized in the basis $\{|a_\alpha\rangle\}_{\alpha=1}^{\alpha=d}$.
}

\begin{equation}\label{f54}
L_a=\sum_\alpha \l_{a\alpha}P_\alpha\;,
\end{equation}
where $\l_{a\alpha}$ are some complex numbers. Moreover, Eq. (\ref{f52}) tells us that the commutation relation of $\mathcal{H}$ and any linear combination of the eigenvectors with purely imaginary eigenvalues, gives another linear combination of such eigenvectors. Again, by using Eq. (\ref{fo1}) and by setting the initial density matrix equal to each projection operator, we obtain

 \begin{equation}\label{f56}
[{\cal H},P_\alpha]=\sum_\beta h_{\alpha\beta} P_\beta\;.
 \end{equation}
One can show that $h_{\alpha \beta}=0$. To see this, multiply Eq. (\ref{f56}) with any projection operator $P_\gamma$ and then take the trace. Therefore, because $\mathcal{H}$ commutes with all $P_\alpha$, we have

 \begin{equation}\label{f57}
 {\cal H}=\sum_\alpha h_\alpha P_\alpha\;,
 \end{equation}
 where $h_\alpha$ are some real numbers.
 
 Now, let us return to Eq. (\ref{f32}) with conditions Eq. (\ref{f57}) and Eq. (\ref{f54}).
Then, the solution of the Lindblad equation would be of the following form 
 
\begin{equation}\label{f58}
\rho(t)=\sum_{\alpha\beta} P_\alpha M P_\beta f_{\alpha\beta}(t)\;,
\end{equation} 
 with the initial conditions $M=\rho_{initial}$ and $f_{\alpha \beta}(0)=1$. The substitution of this solution into Eq. (\ref{f32}) gives
 
 \begin{equation}\label{f59}
 \sum_{\alpha\beta} P_\alpha M P_\beta\frac{d}{dt} f_{\alpha\beta}(t)=-\sum_{\alpha\beta}\lambda_{\alpha\beta} P_\alpha M P_\beta f_{\alpha\beta}(t)\;,
 \end{equation}
 where 
 
 \begin{equation}\label{f60}
 \lambda_{\alpha\beta}= \frac{1}{2}\sum_a\Big| l_{a\alpha}- l_{a\beta}\Big|^2-i\;{\rm Im}\sum_a  l_{n\alpha} l^*_{a\beta}
 +i\Big(h_{\alpha}-h_{\beta}\Big)\;.
 \end{equation}
It is also easy to see that $f_{\alpha \beta}(t)$ has the simple solution

 \begin{equation}\label{f61}
 f_{\alpha\beta}(t)=  e^{-\lambda_{\alpha\beta}t}\;.
 \end{equation}
 Therefore, the density matrix at time $t$ is given by 
 
\begin{equation}\label{f62}
\rho(t)=\sum_{\alpha\beta} P_\alpha (\rho_{initial} )P_\beta e^{-\lambda_{\alpha\beta}t}.
\end{equation} 
From Eq. (\ref{f60}) it is clear that at late time, all terms in the solution Eq. (\ref{f62}) decay exponentially except those that are $l_{a\alpha}= l_{a\beta}$, for all $a$. For the non-degenerate case, we have $l_{a\alpha}= l_{a\beta}$, only if $\alpha = \beta$, and this means the only terms that contribute at late time are $\lambda_{\alpha \alpha}$ which  are of course zero. This leads us to 

\begin{equation}\label{f63}
\lim_{t \rightarrow \infty} \rho(t) = \sum_\alpha P_\alpha\rho(0) P_\alpha = \sum_\alpha \langle \alpha|\rho(0)|\alpha\rangle\  P_\alpha ,
\end{equation}
 which is exactly the Born rule. 
 What about the degenerate case where  $l_{a\alpha}= l_{a\beta}$ even for $\alpha \neq \beta$. To understand this case, it would be necessary to know a little about an \textit{incomplete measurement}. Usually when an actual measurement, done by an experimenter, we do not lead to definite states $|a_\alpha \rangle$ with definite probabilities $p_\alpha$. In fact, often we face  an equivalence classes of non-distinguishable states as the outputs. For example, consider a bipartite system, consisting of two electrons with spins $1/2$. If we only measure the spin of the first electron without disturbing the other, the outputs will fall into two classes $\{|\frac{1}{2}, \frac{1}{2} \rangle,\ |\frac{1}{2}, -\frac{1}{2} \rangle\}$ and $\{|-\frac{1}{2}, \frac{1}{2} \rangle,\ |-\frac{1}{2}, -\frac{1}{2} \rangle\}$. For such measurements, the final density matrix will be
\begin{equation}\label{f64}
\lim_{t \rightarrow \infty} \rho(t) = \sum_C P_C\rho_{initial} P_C ,
\end{equation}
where $P_C$ is an operator which projects every state $|v\rangle$ into the class $C$ and would be defined as 
\begin{equation}\label{f65}
P_C \equiv \sum_{\alpha \in C}P_{\alpha}.
\end{equation}
Apparently, the complete measurement is a special case of incomplete measurement, where each state makes a different class. To describe the incomplete measurement, using of the Lindbald equation, as before,  we need  to have for all $\rho_{initial}$ :
\begin{equation}\label{f66}
\Big[L_a, \sum_C P_C\rho_{initial} P_C \Big]=0.
\end{equation}
By choosing $\rho_{initial}=P_\alpha$ and using the fact that $\sum_{C} P_C P_\alpha P_C = P_\alpha$, we  obtain  the form Eq. (\ref{f54}) again for all $L_a$. If we put Eq. (\ref{f54}) in Eq. (\ref{f66}), then we shall find the following relation 
\begin{equation}\label{f67}
\Big[\sum_\alpha \l_{a\alpha}P_\alpha\ , \sum_C P_C\rho_{initial} P_C \Big]=0= \sum_C \sum_{\beta, \gamma \in C} \Big(l_{a\beta} - l_{a\gamma}\Big) P_{\beta} \rho_{initial} P_{\gamma}.
\end{equation}
 This happens for all initial density matrices if $l_{a\beta} = l_{a\gamma}$, for all $\gamma$ and $\beta$ in the same class. On the same footing, one can show that $h_\alpha = h_\beta$, if $\alpha$ and $\beta$ belong to the same class. Therefore, from Eq. (\ref{f60}), we see that $\lambda_{\alpha \beta}=0$ if $\alpha$ and $\beta$ are in the same class and so $\rho(t)$ in Eq. (\ref{f62}) gives Eq. (\ref{f64}) at late times.
 We see from (\ref{f54}) and (\ref{f57}) that for deriving the Born rule all $L_a$ and $\mathcal{H}$ need to be diagonal with respect to the eigenvectors of whatever is being measured. But it will be not possible to get this if we don't assume the condition Eq. (\ref{f49}). In fact  Eq. (\ref{f49}) is \textit{a necessary and sufficient condition} for the Lindblad equation to give the Born rule. But what does this condition mean? To answer this question, let us do some calculations.\\
 
 %%%%%%%%%%%%%%%%%%%%%%%%%%%%%%%%%%%%%%%%%
 \subsection{Role of the second law in the Born role}
 %%%%%%%%%%%%%%%%%%%%%%%%%%%%%%%%

  Consider the von Neumann entropy $S[\rho(t)]=-{\rm Tr}\Big(\rho(t) \ln\rho(t)\Big)$. This quantity gives us a realization of the concept of entropy in the quantum world. For many of processes known in the classical world, the entropy is a non-decreasing quantity in time. Even in the quantum world, we can find some examples that the entropy of the final state is bigger than the initial state. For example, in the measuring of a density matrix of a pure state (with $S[\rho_{initial}]=0$), the final density matrix turns out to be a mixed state (with $S[\rho_{final}] \geq 0$). Therefore, it would be natural to ask, under what circumstances does the von Neumann entropy never decrease? To find this condition we have to find the time derivative of the von Neumann entropy. First, we note that for a functional $F[\rho(t)] $ we have

  \begin{equation}\label{f68}
  \frac{d}{dt}{\rm Tr}\ F[\rho(t)] = {\rm Tr}\Big( \frac{d F[\rho(t)]}{d\rho(t)} \frac{d\rho(t)}{dt} \Big).
  \end{equation} 
To see this, supposed   $ | \Psi_a \rangle$ be time dependent eigenvectors of $\rho(t)$, with eigenvalues $p_a$. Then, for LHS of Eq. (\ref{f68}), one finds

 \begin{equation}\label{f69}
 \frac{d}{dt}{\rm Tr}\ F[\rho(t)] =  \frac{d}{dt}\left( \sum_a \langle \Psi_a |\ F[\rho(t)]\ | \Psi_a \rangle \right) = \frac{d}{dt} \sum_a F[p_a(t)] = \sum_a F^{\prime}[p_a(t)] \frac{dp_a(t)}{dt}.
 \end{equation} 
 On the other hand, one can show that:
 
 \begin{equation}\label{f70}
 \langle\Psi_a |\frac{d\rho(t)}{dt} | \Psi_a \rangle = \frac{d}{dt}\Big( \langle\Psi_a |\rho(t) | \Psi_a \rangle \Big) - \frac{d}{dt}\Big( \langle\Psi_a |\Big)\ \rho(t) | \Psi_a \rangle - \langle\Psi_a | \rho(t)\ \Big( \frac{d}{dt} | \Psi_a \rangle \Big) = \frac{dp_a(t)}{dt}.
 \end{equation} 
 By substituting Eq. (\ref{f70}) in Eq. (\ref{f69}) and by noting that

 \begin{equation}\label{f71}
 \sum_a F^{\prime}[p_a(t)] \frac{dp_a(t)}{dt} = {\rm Tr}\Big( \frac{d F[\rho(t)]}{d\rho(t)} \frac{d\rho(t)}{dt} \Big).
 \end{equation} 
 One can gets Eq. (\ref{f68}). Thus, for the time derivative of the von Neumann entropy we get

 \begin{equation}\label{f72}
 \frac{d}{dt} \mathcal{S}[\rho] = - {\rm Tr}\left(\frac{d\rho(t)}{dt} \ln \rho  \right).
 \end{equation} 
If one uses the Lindblad equation (\ref{f32}) and the fact that 

 \begin{equation}\label{f73}
 {\rm Tr}\Big(\left[\mathcal{H}, \rho\right]  \ln \rho  \Big) = {\rm Tr}\Big(  \mathcal{H} \left[\rho , \ln \rho \right] \Big) = 0.
 \end{equation} 
 
The time derivative of the von Neumann entropy finds the following form 

 \begin{equation}\label{f74}
 \frac{d}{dt} \mathcal{S}[\rho] = \sum_a {\rm Tr}\Big[ L^{\dagger}_{a} L_a \rho \ln \rho \Big] - \sum_a {\rm Tr}\Big[  L_a \rho L^{\dagger}_{a}\ln \rho \Big] = \sum_{ij,a} {|(L_a)_{ij}|}^2 p_j \Big(\ln p_j - \ln p_i \Big).
 \end{equation} 
 
To make further progress, we shall use an equality:
 \begin{equation}\label{f78}
 \frac{1}{x} + \ln x \geqslant 1 \ \ \ x > 0,
 \end{equation} 
 where, we shall have the equality  if $x=1$. Now, setting $ x= \frac{p_j}{p_i}$ in this inequality gives

 \begin{equation}\label{f79}
 p_j \Big( \ln p_j - \ln p_i \Big) \geqslant p_j - p_i.
 \end{equation} 
 Let us use Eq. (\ref{f79}) in Eq. (\ref{f74}) to find the following inequality 
 
 \begin{equation}\label{f80}
 \frac{d}{dt} \mathcal{S}[\rho] = \sum_{ij,a} {|(L_a)_{ij}|}^2 p_j \Big(\ln p_j - \ln p_i \Big) \geqslant \sum_{ij,a} {|(L_a)_{ij}|}^2  p_j - \sum_{ij,a} {|(L_a)_{ij}|}^2   p_i = \sum_{ij,a} \left[{|(L_a)_{ij}|}^2 - {|(L_a)_{ji}|}^2 \right]  p_j .
 \end{equation} 
 One can see that the necessary and  sufficient  condition for the von Neumann entropy to be non-decreasing, is that :

 \begin{equation}\label{f81}
 \sum_{i,a} \left[{|(L_a)_{ij}|}^2 - {|(L_a)_{ji}|}^2 \right] \geqslant 0 \ \ \ \Rightarrow\  \sum_a L_a^{\dagger} L_a \geqslant \sum_a L_a L_a^{\dagger} .
 \end{equation} 

Note that when we limit the Hilbert space to a finite dimensional space, the inequality Eq. (\ref{f81}) turns out to be equality. To see this, suppose there is at least one $j$ in Eq. (\ref{f81}) for which $\sum_{i,a} {|(L_a)_{ij}|}^2 > \sum_{i,a} {|(L_a)_{ji}|}^2$. Then by summing over all $j$, one gets $\sum_{ij,a} {|(L_a)_{ij}|}^2 > \sum_{ij,a} {|(L_a)_{ji}|}^2$, which is not possible. Thus the relation (\ref{f81}) should be an equality. It is interesting that in the case of CSL models the condition Eq. (\ref{f49}) is automatically satisfied, because it turns out that all $L_a$ in such models are Hermitian. There is a nice review on CSL models \cite{Bassi:2012bg} and we encourage the reader to study it, if he or she is not familiar with this subject.

The Lindblad equation not only describes many sorts of open systems but also it may appear in some modified versions of quantum mechanics. In such modified quantum theories, the first term in LHS of Eq. (\ref{f32}) defines the standard quantum mechanics and $\mathcal{H}$ has the role of the Hamiltonian and the rest terms turns out to be corrections to the standard theory. Here, a question immediately arises: \textit{How can one observe the effects of these corrections if they exist?} 
To see how these corrections can be observed, we shall use the Ramsey interferometer. In the next section, we study the Ramsey interferometer mechanism which is the basis of the Atomic Clocks.\\

%%%%%%%%%%%%%%%%%%%%%%%%%%%%%%%%%%%%%%%%%%%%%%%%%%%%%%%%%%%%
\section{Correction in Ramsey Interferometers}
%%%%%%%%%%%%%%%%%%%%%%%%%%%%%%%%%%%%%%%%%%%%%%%%%%%%%%%%%%%%%%
 In 1949, Norman Ramsey developed a method which allows extremely accurate measurements of molecular or atomic transition frequencies. In atoms or molecules with discrete bound states of energies $E_m$, we are interested in the probabilities of transitions between two states during the time $t$, when we expose a perturbation to the system. Let's first review Ramsey's work \cite{Ramsey} in ordinary quantum mechanics and then we shall return to the above question. \footnote{Here, we will follow Weinberg \cite{Weinberg-clocks} with a little different way.}

Suppose a system with the Hamiltonian 
 
 \begin{equation}\label{ff79}
H(t)= H_0 + H^{\prime}(t),
 \end{equation}
 where $H_0$ is the time-independent free Hamiltonian and $ H^{\prime}(t)$ is a small time-dependent perturbation, which depends on some external fields. The evolution of the density matrix in standard quantum mechanics is given by 
 \begin{equation}\label{ff80}
 \partial_t \rho(t) = -i \left[H(t), \rho(t)\right] .
 \end{equation}
This equation has the following solution

 \begin{equation}\label{ff81}
 \rho(t) =  \sum_{m,n} f_{mn}(t)\ e^{-i(E_m - E_n) t}\  |m \rangle  \langle n|,
 \end{equation}
where vectors $|m\rangle$ are orthonormal eigenvectors of $H_0$, with eigenvalues $E_m$, and $f_{ij}(t)$ are components of a Hermitian matrix i.e. $f_{ij}(t)=f^{*}_{ji}(t)$. We also suppose eigenvectors $|m\rangle$ to be stable states. If we put this solution into Eq. (\ref{ff80}) we shall get

 \begin{equation}\label{ff82}
 i \sum_{m,n} \dot{f}_{mn}(t)\ e^{-i(E_m - E_n) t}\  |m \rangle  \langle n| = \sum_{m,n} f_{mn}(t)\ e^{-i(E_m - E_n) t}\Big(  H^{\prime}(t) |m \rangle  \langle n| - |m \rangle  \langle n| H^{\prime}(t) \Big).
 \end{equation}
 By multiplying this equation into $\langle i |$ from left and into $|j\rangle$ from right, we shall find 
 
 \begin{equation}\label{ff83}
 i  \dot{f}_{ij}(t)\ e^{-i(E_i - E_j) t}\   = \sum_{m}\left( H_{im}^{\prime}(t) f_{mj}(t)\ e^{-i(E_m - E_j) t}    - H_{mj}^{\prime}(t) f_{im}(t)\ e^{-i(E_i - E_m) t}   \right),
 \end{equation}
 
 where
 
 \begin{equation}\label{ff84}
H_{ij}^{\prime}(t) \equiv \langle i| H^{\prime}(t) |j \rangle.
 \end{equation}
 
From Eq. (\ref{ff83}), it would be easy to see that the coefficients $f_{ij}(t)$ satisfy the following differential equations 

\begin{equation}\label{ff85}
i  \dot{f}_{ij}(t)  = \sum_{m}\left( H_{im}^{\prime}(t) f_{mj}(t)\ e^{-i(E_m - E_i) t}    - H_{mj}^{\prime}(t) f_{im}(t)\ e^{-i(E_j - E_m) t}   \right).
\end{equation} 
 
In the Ramsey interferometer, the perturbation $H^{\prime}(t)$ is supposed to be \textit{monochromatic}. We say a perturbation $H^{\prime}(t)$ is monochromatic if it oscillates with a single frequency $\omega$ and its dependence on time is of the form 
\begin{equation}\label{ff86}
H^{\prime}(t) = - U e^{-i\omega t} - U^{\dagger} e^{i\omega t},
\end{equation} 
where $U$ is a non-singular matrix. For example, consider a hydrogen atom in its ground state and suppose we have exposed it to an electromagnetic wave with a plane polarization. If the wavelength of this wave is much larger than Bohr's radius, then this leads us to a monochromatic perturbation in the Hamiltonian\footnote{In such cases, we consider only the electric field. The magnetic component of the electromagnetic wave has no contribution to the Hamiltonian because the magnetic force for a non-relativistic electron placed in the electromagnetic field, is less than the electric force by a factor of order $v/c$, where is the velocity of the electron.}. If we use Eq. (\ref{ff86}) in Eq. (\ref{ff85}), we obtain  

\begin{eqnarray}\label{ff87}
i  \dot{f}_{ij}(t)  &=& -\sum_{m}\left( U_{im} f_{mj}(t)\ e^{-i(\omega + E_m - E_i ) t}    + U^{*}_{mi} f_{mj}(t)\ e^{-i( E_m -\omega  - E_i) t}   \right)\\ \nonumber
&+& \sum_{m}\left( U_{mj} f_{im}(t)\ e^{-i(\omega + E_j - E_m ) t}    + U^{*}_{jm} f_{im}(t)\ e^{-i(E_j - E_m -\omega  ) t}   \right).
\end{eqnarray}

 Now, suppose the perturbation frequency $\omega$ is tuned out to be close to one of the resonance frequencies $(E_e-E_g)$ where $E_e$ and $E_g$ are energies of the ground state and the excited state respectively ($E_e > E_g$). Ignoring all terms in Eq. (\ref{ff87}) with coefficients that oscillate rapidly, and keeping the terms with relatively small oscillation frequency $\pm \Big( \omega - (E_e-E_g)\Big) $, one gets the following system of differential equations  
 
\begin{equation}\label{ff88}
i  \dot{f}_{ee}(t)  =  U^{*}_{eg} f_{eg}(t)\ e^{i \Delta \omega t}  -  U_{eg} f_{ge}(t)\ e^{- i \Delta \omega t},
\end{equation} 
 \begin{equation}\label{ff89}
 i  \dot{f}_{gg}(t)  = - U^{*}_{eg} f_{eg}(t)\ e^{i \Delta \omega t}  -  U_{eg} f_{eg}(t)\ e^{- i \Delta \omega t}
 \end{equation} 
 \begin{equation}\label{ff90}
 i  \dot{f}_{eg}(t)  =  U_{eg} f_{ee}(t)\ e^{-i \Delta \omega t}  -  U_{eg} f_{gg}(t)\ e^{- i \Delta \omega t},
 \end{equation} 
 where 
 \begin{equation}\label{ff91}
 \Delta \omega \equiv \omega - (E_e - E_g).
 \end{equation}
 
One can use a Laplace transformation or a similar transformation (to diagonalize the matrix of coefficients) to solve this system of differential equations. The solutions are  

 \begin{equation}\label{ff92}
f_{gg}(t) = \frac{1}{2} R^2 \left[ (1+A^2)\Big\{\Omega^2 +\frac{\Delta\omega^2}{4}\Big\} + {|U_{eg}|}^2 \cos(2\Omega t +2B)\right], 
 \end{equation}
 \begin{equation}\label{ff93}
 f_{ee}(t) = \frac{1}{2}R^2{|U_{eg}|}^2  \Big[ (1+A^2) -\cos(2\Omega t +2B)\Big] ,
 \end{equation}
 \begin{equation}\label{ff94}
 f_{ge}(t) = -\frac{1}{2}R^2 e^{i \Delta \omega t} U^{*}_{eg}  \left[ D - \frac{\Delta\omega}{2}\cos(2\Omega t +2B) +i\Omega \sin(2\Omega t +2B) \right] ,
 \end{equation}
 
where
 \begin{equation}\label{ff95}
 \Omega^2 = \frac{\Delta\omega^2}{4} + {|U_{eg}|}^2,
 \end{equation}
 
and $A, B, D, R$ are some real constants. We note that the number of the real constants is equal to the number the real parameters of the density matrix (in the two dimensional space defined by the excited state $|e\rangle$ and the ground state $|g\rangle$). Therefore, they are determined by the initial constants $f_{ij}(0)$. For example if we have an ensemble of the same atoms  all of which are in the ground state $|g\rangle$, at time $t=0$, then  
 \begin{equation}\label{ff96}
 f_{ij}(0) = \delta_{ig} \delta_{jg}.
 \end{equation} 

This leads to the following values for integration constants:
\begin{equation}
A=B=0,\ \ R^2 = \frac{1}{\Omega^2},\ \ D=\frac{\Delta \omega}{2}.
\end{equation}

With these values for constants the solutions in Eq. (\ref{ff92}), Eq. (\ref{ff94}) and Eq. (\ref{ff93})  will be simplified :
 \begin{equation}\label{ff98}
 f_{gg}(t) = \cos^2(\Omega t) + \frac{\Delta \omega^2}{4 \Omega^2} \sin^2(\Omega t),
 \end{equation} 
 \begin{equation}\label{ff99}
 f_{ee}(t) = \frac{{|U_{eg}|}^2}{\Omega^2} \sin^2(\Omega t),
 \end{equation} 
\begin{equation}\label{ff100}
f_{eg}(t) = \frac{i U_{eg}}{2 \Omega} e^{-i \Delta \omega t} \Big[ \sin(2\Omega t) + \frac{i\Delta \omega}{ \Omega} \sin^2(\Omega t) \Big].
\end{equation} 
 
In the Ramsey interferometer, the atoms (or molecules) in the ground state are exposed to a pulse of microwave radiation for a short time $t_1$. They then travel without any interaction with the external field  for a much longer time $T$, and then they are again exposed to the pulse of microwave radiation for another short time $t_2$ and finally go outside the interferometer to a detector that counts atoms in the ground state or in the excited state. The advantage of Ramsey's trick is that the probabilities of finding the atoms in the excited state $|e\rangle$ are very sharply peaked at $\Delta\omega=0$. Therefore, one would be able to make a very accurate measurement of the resonance frequency $(E_e-E_g)$ by tuning the frequency of the microwave radiation pulse $\omega$. To find the transition probability  in the Ramsey interferometer, we have to divide the density matrix or equivalently the time-dependent coefficients $f_{ij}(t)$ into the three parts. In the first part we suppose atoms in the ground state  $|g\rangle$ to start at $t=0$ their travel in the microwave radiation pulse of frequency $\omega$, and at $t=t_1$ finish the first part of this trip. We call $f_{ij}$ during $[0,t_1]$ as $f^{(1)}_{ij}$. In the second part, the atoms freely travel from $t=t_1$ to $t=t_1 + T$. Because during this time there is no interaction with the external fields, the coefficients $f_{ij}(t)$ turn out to be constants. If we call $f_{ij}$ during $[0,t_1]$ as $f^{(2)}_{ij}$, we shall have $f^{(2)}_{ij}(t)=f^{(1)}_{ij}(t_1)$ for the second part. In the last part, the atoms again enter  the external field up to time $t=t_1+t_2+T$. We label $f_{ij}$ in the last part as $f^{(3)}_{ij}$. For the first part $f^{(1)}_{ij}(t)$ are what we have found previously i.e. Eqs. (\ref{ff98}), (\ref{ff99}) and (\ref{ff100}). But $f^{(3)}_{ij}(t)$ in the third part have the solutions Eqs. (\ref{ff92}), (\ref{ff93}) and (\ref{ff94}), where we should determine $A, B, D, R$. To find these constants it will be enough to use the boundary conditions 
 \begin{equation}\label{ff101}
f^{(3)}_{ij}(t_1 + T) =  f^{(1)}_{ij}(t_1) .
 \end{equation} 
 Using the results of these boundary conditions, one can find $f^{(3)}_{ij}$ at the end of third part. That is
 
 \begin{equation}\label{ff102}
 f^{(3)}_{ee}(t_1 +t_2+ T) = \sin^2(\Omega t)\  {\Big| e^{-i \Delta \omega T} \sin(\Omega t_2) \cot(\Omega t_t) + \cos(\Omega t_2)\Big| }^2.
 \end{equation} 
 
 In fact, it gives the transition probability ${Pb}_e \equiv {\rm Tr}(\rho |e\rangle\langle e |)$. Usually to simplify this result, people  set $t_1=t_2=\tau$. This can always be done, if we construct the interferometer in such a way that the length of the traveling path for atoms at the first part would be the same as the  length of the traveling path at the last part. Under this condition, we have 
 \begin{equation}\label{ff103}
{Pb}_e =  f^{(3)}_{ee}(2\tau+ T) = \sin^2 (2\Omega \tau)\  { \cos^2(\frac{\Delta \omega T}{2})} =\frac{1}{2} \sin^2 (2\Omega \tau) \Big[ 1+ \cos\Big(\Delta \omega T\Big)\Big].
 \end{equation} 
One can easily see from this relation that the transition probability is very sharp at $\Delta\omega=0$. We should be careful that there is always some spread in the velocity of different atoms. Suppose that because of a spread in velocities, the probability distribution that an atom spends a time between $T$ and $T+dT$ (during its trip in the second part) is Gaussian

 \begin{equation}\label{ff104}
 P(T) = \frac{1}{\sqrt{\pi \sigma^2} }\exp \left[ -\frac{1}{\sigma^2}{\Big( T- T_0 \Big)}^2 \right],
 \end{equation} 
 where $T_0$ is the mean time between pulses and $\sigma$ is the spread in $T$. Now one can easily find the fraction of atoms in the excited state. This would be  
 
 \begin{equation}\label{ff105}
 < {Pb}_e > = \int_{-\infty}^{+\infty} dT\ P(T)\ {Pb}_e  = \frac{1}{2}\sin^2 (2\Omega \tau) \left[ 1 +\cos\Big(\Delta\omega T_0\Big) \exp \Big( -\frac{1}{4}{\Delta\omega}^2 {\sigma^2}\Big) \right].
 \end{equation} 
 
\begin{figure}[h]
    \begin{center}
        \includegraphics[scale = 0.74]{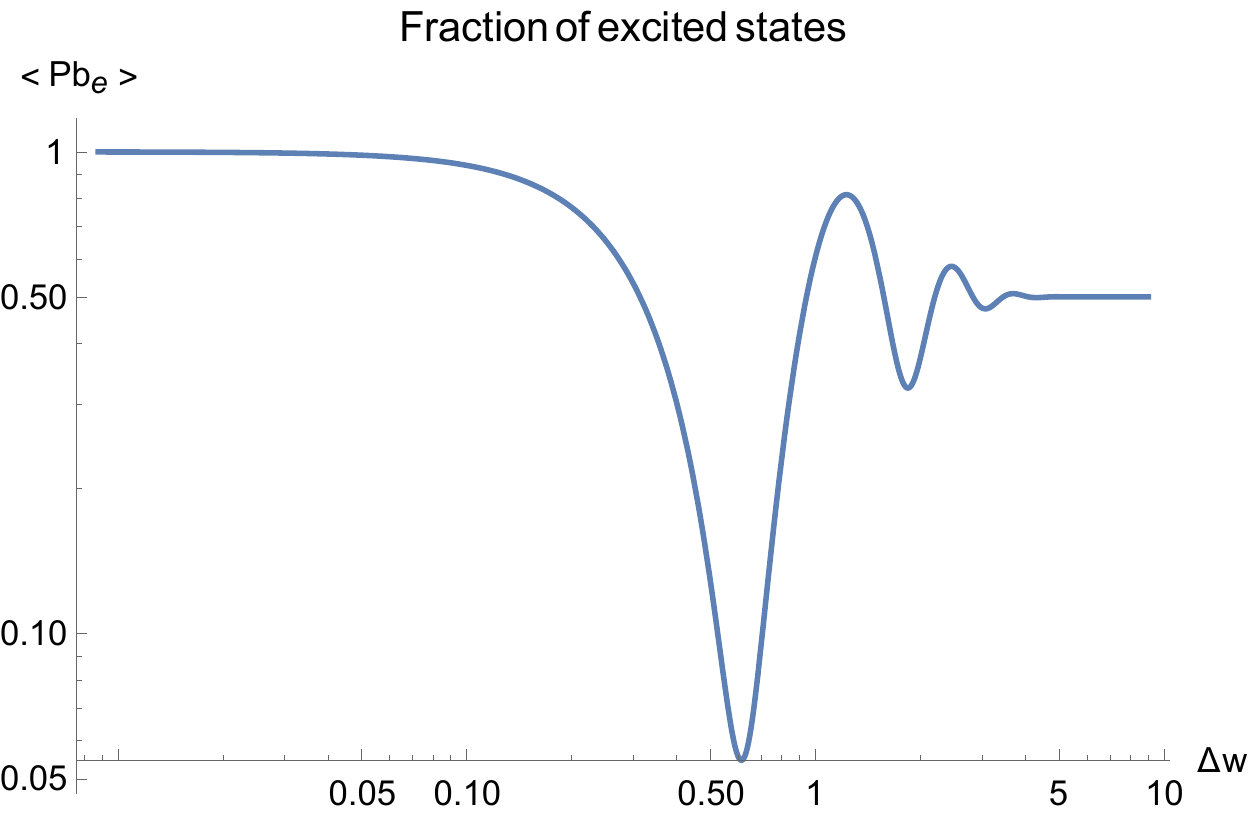}
        \hspace*{10mm} \caption{ \label{standard}
            The diagram of the fraction of atoms in the excited state in the standard quantum mechanics. }
    \end{center}
\end{figure}

\begin{figure}[h]
    \begin{center}
        \includegraphics[scale = 0.74]{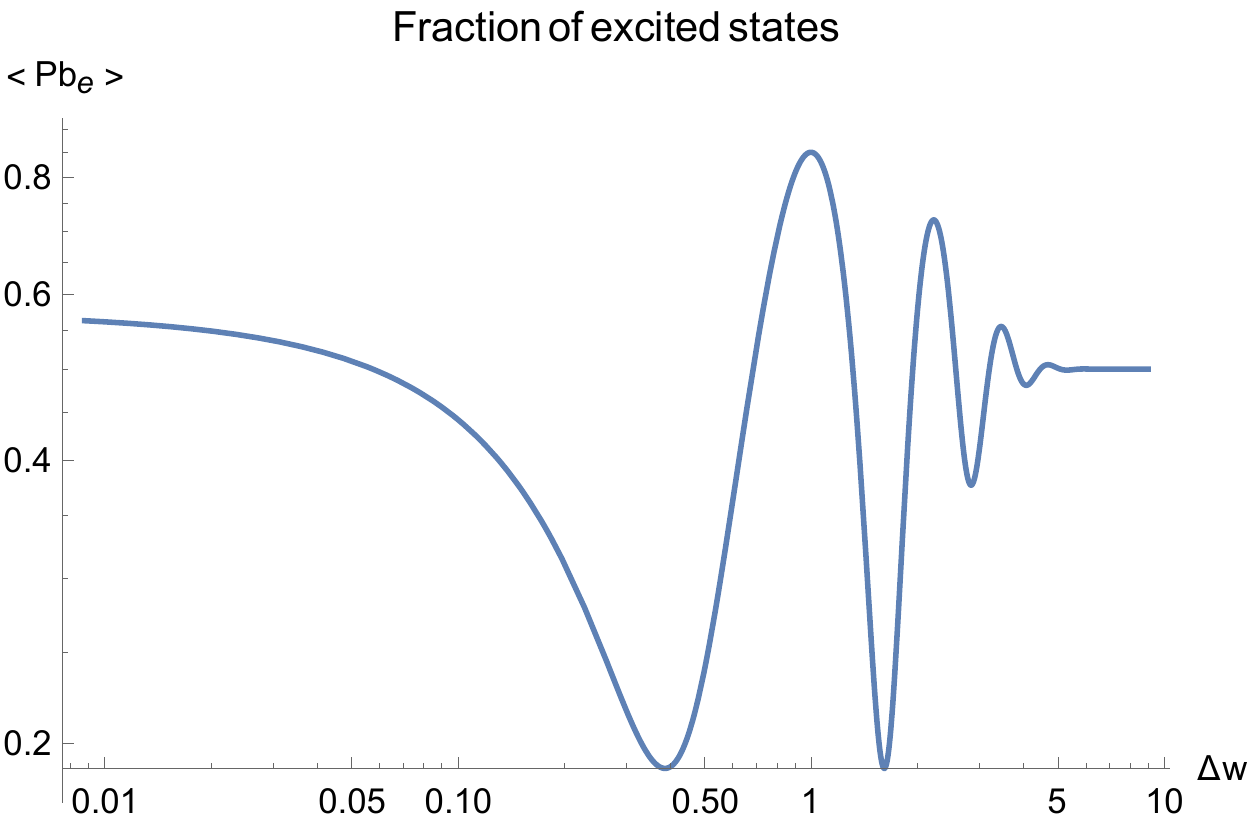}
        \hspace*{10mm} \caption{ \label{correction}
        The diagram of  the fraction of atoms in the excited state in the linearly modified quantum theories. This diagram approaches the standard diagram when $\lambda_{\alpha \beta}\rightarrow 0$}
    \end{center}
\end{figure}

 Now, let us go back to Eq. (\ref{f32}) and solve it for Ramsey interferometer. We assume that the von Neumann entropy is non-decreasing for the modified theory Eq. (\ref{f32}), and so for the corrections, we have the condition (\ref{f49}). Because we have assumed the vectors $|m\rangle$ are stable\footnote{This happens when the rate of radiative transitions are very small and can be ignored.}, one can easily check the fact that the excited state $|e\rangle$ and the ground state $|g\rangle$ are eigenstates of $\mathcal{H}$ and $L_a$ in Eq. (\ref{f32}). To see this, set $\rho(t)=|e\rangle\langle e|$ and multiply this equation on the left with projection $P_e=|e\rangle\langle e|$. By taking the trace we find that
 \begin{equation}\label{ff106}
     0={\rm Tr}\left(\sum_a [L_a,P_m]^\dagger[L_a,P_m]\right)
     +{\rm Tr}\left(P_m\sum_a \Big(L_a^\dagger L_a-L_a L_a^\dagger\Big)\right)\;,
 \end{equation}
 where we have used the fact that for stable states $\dot{\rho}=0$ and ${\rm Tr}\Big(\rho [\mathcal{H}, \rho] \Big) = 0$. Regarding the condition (\ref{f49}), the second term on RHS vanishes and thus we have :
 \begin{equation}\label{ff107}
[L_a, P_e]=[L^{\dagger}_a, P_e]=0 .
 \end{equation} 
 Because the stable density matrix $\rho=P_e$ commutes with all $L_a$ and $L^{\dagger}_a$, if we put it in Eq. (\ref{f32}), we easily find
 \begin{equation}\label{ff108}
 [\mathcal{H}, P_e]=0.
 \end{equation}
 If the commutators Eq. (\ref{ff107}) and Eq. (\ref{ff108}) act on $|e\rangle$, one can easily see that the state $|e\rangle$ is an eigenstate of $\mathcal{H}$ and $L_a$. The same argument works for the ground state  and so $|g\rangle$ is also an eigenstate of $\mathcal{H}$ and $L_a$.
 With this preliminary results, at least in the two dimensional space $\Big\{ |e\rangle, |g\rangle\Big\}$, Eq. (\ref{f32}) has the solution Eq. (\ref{f62}), where the indices $\alpha, \beta$ change between the two states $e, g$.
 Now, we shall study the effects of the corrections on the transition probability, derived by Ramsey's trick. We assume that the exposure time $t_1=t_2=\tau$ in the first and the third parts of the interferometer is short enough so that $|\tau \lambda_{\alpha \beta}|\ll 1$ and therefore we ignore the corrections when we are exposing the external field on the atoms. This assumption tells us that the solutions for the first and the last part of the interferometer would be the same as before. But in the second part of the interferometer,  we have the following solution for the density matrix
 
 \begin{equation}\label{ff109}
 \rho(t)=\sum_{i,j= e, g} P_i (\rho_{initial} )P_j e^{-\lambda_{ij}t}= \sum_{i,j= e, g} \tilde{f}_{ij}(t) e^{-(E_i-E_j)t} |i\rangle\langle j|,
 \end{equation}
 where in the first equality $\lambda_{ij}$ are given by Eq. (\ref{f60}) and in the second equality we have separated the phases which depend on the eigenvalues of $\mathcal{H}=H_0$\footnote{Here we just set $h_i =E_i$ where $h_i$ are defined by (\ref{f57})}. We note that 
 \begin{equation}\label{ff110}
\tilde{f}_{ij}(t) \propto e^{-\tilde{\lambda}_{ij}t},
 \end{equation}
 where 
 \begin{equation}\label{ff111}
 \tilde{\lambda}_{ij}\equiv \frac{1}{2}\sum_a\Big| l_{a i}- l_{a j}\Big|^2-i\;{\rm Im}\sum_a  l_{n i} l^*_{a j}.
 \end{equation}
We saw that in the standard theory the coefficients $f_{ij}$ in Eq. (\ref{ff81}) were constant in the second part of the interferometer, but using Eq. (\ref{ff109}) and Eq. (\ref{ff110}) one immediately finds out that the coefficients of the density matrix, apart from the energy phases $e^{-(E_i-E_j)t}$, are no longer time independent in the second part and by using the boundary conditions, they would find the time-dependent solution
\begin{equation}\label{ff-112-1}
f^{(2)}_{ij}(t)=f^{(1)}_{ij}(t_1) e^{-\tilde{\lambda}_{ij}(t-t_1)} .
\end{equation} 
This would change the boundary conditions Eq. (\ref{ff101}) as
\begin{equation}\label{ff112}
f^{(3)}_{ij}(t_1 + T) =  f^{(1)}_{ij}(t_1)\ e^{-\tilde{\lambda}_{ij}T}.
\end{equation} 
 Now, let us find the transition probability $Pb_e$ again. For simplicity choose the amplitude of the external field to be much bigger than $\Delta\omega$ or in other words $U_{eg} \gg \Delta\omega$ and hence $\Omega \approx U_{eg}$. Using Eqs. (\ref{ff92})-(\ref{ff100}) and Eq. (\ref{ff-112-1}),  one will find that
 \begin{eqnarray}
 && f_{ee}(\tau)=\sin^2(\Omega\tau)\;,~~~~~ f_{gg}(\tau)=\cos^2(\Omega\tau)\;,\nonumber\\
 && f_{eg}(\tau)=ie^{-i\Delta\omega\tau}\cos(\Omega \tau)\ \sin(\Omega\tau)\;,
 \end{eqnarray}
 \begin{eqnarray}
 && f_{ee}(\tau + T)=\sin^2(\Omega\tau)\;,~~~~~ f_{gg}(\tau+ T)=\cos^2(\Omega\tau)\;,\nonumber\\
 && f_{eg}(\tau+ T)=ie^{-i\Delta\omega\tau} e^{-\lambda_{eg}T}\cos(\Omega \tau)\ \sin(\Omega\tau)\;.
 \end{eqnarray}
 Using the boundary conditions Eq. (\ref{ff112}), we get the following relation for transition probability
 
 \begin{equation}\label{ff116}
 {Pb}_e=f_{ee}(2\tau+T)=\frac{1}{2}\sin^2(2\Omega\tau)\left[
 1+e^{-({\rm Re}\tilde{\lambda}_{eg}) T}\cos\Bigg(\Big(\Delta\omega-{{\rm Im}\tilde{\lambda}_{eg}}\Big)T\Bigg)\right]\;.
 \end{equation}
 Finally, we should compute what one sees in the laboratory .i.e. the fraction of atoms in the excited state. Using the Gaussian distribution Eq. (\ref{ff104}), we get 
 \begin{eqnarray}\label{fff117}
 & &< {Pb}_e > = \int_{-\infty}^{+\infty} dT\ P(T)\ {Pb}_e 
  = \frac{1}{2}\sin^2 (2\Omega \tau) \\ \nonumber & &\left[ 1 +e^{-{\rm Re}\tilde{\lambda}_{eg} \Big(T_0 -{\rm Re}\tilde{\lambda}_{eg}\frac{\sigma^2}{4}\Big)}\cos\Big((\Delta\omega -{\rm Im}\tilde{\lambda}_{eg} ) (T_0-{\rm Re}\tilde{\lambda}_{eg}\frac{\sigma^2}{2})\Big) \exp \Big( -\frac{1}{4}{(\Delta\omega -{\rm Im}\tilde{\lambda}_{eg})}^2 {\sigma^2}\Big) \right] \label{ff117}.
 \end{eqnarray}
There are two significant differences between Eq. (\ref{fff117}) and Eq. (\ref{ff105}). The first is that the exponentially damping factor $\exp\Big({-{\rm Re}\tilde{\lambda}_{eg} (T_0 -{\rm Re}\tilde{\lambda}_{eg}\frac{\sigma^2}{4})}\Big)$ in Eq. (\ref{fff117})  arises because of the real part of $\tilde{\lambda}_{eg}$. The second is that the shifts in the cosine and the exponent function arise because of  both imaginary and real parts of $\tilde{\lambda}_{eg}$. We expect if there is any linear connection to the standard theory, it can be observed by the study of the outputs in a Ramsey interferometer (See Fig.\ref{standard} and Fig.\ref{correction} )        \\

 As mentioned before, Eq. (\ref{f9}) and consequently, the Lindblad equation may arise in different modified quantum theories or open quantum systems. One might be interested in those theories of modified quantum mechanics or open quantum systems in which the state vector undergoes a stochastic evolution (very fast for macroscopic systems and slow for microscopic systems). Some collapse models and all of the CSL models can be included in such a study. In such theories, we are able to attribute a probability density $\mathcal{P}(\psi, t)$ to the system under study, where $|\psi\rangle$ is the time-independent state vector of system(in Heisenberg picture) and  $\mathcal{P}(\psi, t) d\psi$ is the probability for the wave function of the system to be in a small volume $d\psi$ around $|\psi\rangle$. The volume element $d\psi$ is defined on a complex manifold in which each point of the manifold is a normalized vector in the physical Hilbert space. This volume element should not depend on the chose of the basis for Hilbert space, and hence it must be invariant under all unitary transformations. This is defined as
 \begin{equation}\label{ff118}
 d\psi\equiv \delta\left(1-\sum_i|\psi_i|^2\right) d\psi_1 \wedge d\psi_1^{*} \wedge d\psi_1 \wedge d\psi_1^{*} \wedge \ldots  =\delta\left(1-\sum_i|\psi_i|^2\right)\prod_i d|\psi_i|^2\;\frac{d{\theta_i}}{2\pi}\, ,
 \end{equation}
 
 where $\psi_i$ are the components of state vector $|\psi\rangle$ in an arbitrary orthonormal basis, and $\theta_i$ are arguments of $\psi_i =|\psi_i| e^{i\theta_i} $. The delta function appears here because of invariant norm condition $\sum_i|\psi_i|^2=1$. Now, we define the function $\Pi(\psi,\psi^{\prime}; t-t^{\prime})$, as the probability density of the system to be in the state $|\psi\rangle$ at time  $t> t^{\prime}$, if the wave function of the system at the earlier time $t^{\prime}$ is $|\psi^{\prime}\rangle$. One notes that this quantity is a function of the elapsed time $t-t^{\prime}$, because we have applied the time-translation invariance assumption. From principles of probability theory it is clear that if the wave function $|\psi^{\prime}\rangle$ has probability density $\mathcal{P}(\psi^{\prime},t^{\prime})$ at the time $t$, then at time $t> t^{\prime}$ the probability density for $|\psi\rangle$ will be
 \begin{equation}\label{ff119}
 \mathcal{P}(\psi,t)=\int d\psi'\;\Pi(\psi,\psi';t-t')\,\mathcal{P}(\psi',t')\;.
 \end{equation}
  In the theories which we are talking about, $\Pi(\psi,\psi';t-t')$ has an important property : The space of bilinear functions of $\psi$ is invariant under the action of $\Pi(\psi,\psi';t-t')$, i.e.
  \begin{equation}\label{fff120}
  \int d\psi\; \Pi(\psi,\psi';\tau) \psi_i\psi_j^* =\sum_{i'j'} K_{ii',jj'}(\tau) \psi_i'\psi_j'^*.
  \end{equation}
  This happens because in such theories(for example in case of CSL models), the wave function of the system at time $t$ is given by $|\psi,t\rangle$, where the components of this vector change stochastically with time (usually exponentially), and hence we have for the probability density $\Pi(\psi,\psi';t-t')=| \langle \psi,t|\psi',t'\rangle|^2$. 
   One can go further and find a first order differential equation by differentiating both sides of Eq. (\ref{ff119}) with respect to $t$ at $t=t'$. The fixed points of the resulting differential equation have a crucial role in the study of the Born rule. However, we shall not study this differential equation and encourage the reader to see \cite{coll}. 
 Now, let us describe why such theories should obey relation Eq. (\ref{f9}) and consequently Eq. (\ref{f32}) for the dynamics of density matrices. To define the density matrix in such theories we note that $\mathcal{P}(\psi,t) d\psi$ is the weight of the state $|\psi\rangle\langle \psi |$. Thus, we have
 \begin{equation}
 \rho(t)\equiv \int d\psi\;\mathcal{P}(\psi,t)|\psi\rangle\langle\psi|,
 \end{equation}
or in terms of components of the density matrix
 \begin{equation}\label{ff121}
 \rho_{ij}(t)\equiv \int d\psi\;\mathcal{P}(\psi,t)\psi_i\psi_j^*\;,
 \end{equation}
 where $\psi_i = \langle i|\psi\rangle$. If we put Eq. (\ref{ff119}) into Eq. (\ref{ff121}) we shall find 
  \begin{equation}\label{ff122}
 \rho_{ij}(t)\equiv \int d\psi\;\mathcal{P}(\psi,t)\psi_i\psi_j^* = \int d\psi' \mathcal{P}(\psi',t') \int d\psi\; \Pi(\psi,\psi';t-t') \psi_i\psi_j^*\, .
 \end{equation}
 Now if we insert Eq. (\ref{fff120}) into Eq. (\ref{ff122}), we get
\begin{equation}\label{ff122}
\rho_{ij}(t)\equiv \int d\psi\;\mathcal{P}(\psi,t)\psi_i\psi_j^* = \int d\psi' \mathcal{P}(\psi',t') \sum_{i'j'}K_{ii',jj'}(t-t') \psi_i'\psi_j'^* =\sum_{i'j'} K_{ii',jj'}(t-t') \rho_{i'j'}(t'),
\end{equation} 
which is what was considered in Eq. (\ref{f9}). \\

%%%%%%%%%%%%%%%%%%%%%%%%%%%%%%%%%%%%%%%%%%%%%%%%%%
%\section{Conclusion}
%%%%%%%%%%%%%%%%%%%%%%%%%%%%%%%%%%%%%%%%%%%%%%%%%%
%%    In this paper, in terms of the basic concept of the density matrix in an open quantum system and modification of the quantum mechanics, we derived  Kossakowski-Lindblad equation and different properties of this equation are reviewed. We surveyed the conditions that the Lindblad equation leads to Born role and its connection with the von Neumann entropy.  Next, we used a pedagogical approach to present an atomic clocks mechanism for an open quantum system and modification of the quantum mechanics...\\

%%%%%%%%%%%%%%%%%%%%%%%%%%%%%%%%%%%%%%%%%%%%%%%%%%%%%%%%
%{\bf Acknowledgments:}

%We would like to thank ..... for useful comments and discussion.\\

\clearpage
%%%%%%%%%%%%%%%%%%%%%%%%%%%%%%%%%%%%%%%%%%%%%%%%%%%%%%%%
\appendix
%%%%%%%%%%%%%%%%%%%%%%%%%%%%%%%%%%%%%%%%%%%%%%%%%%%%%%%%

%%%%%%%%%%%%%%%%%%%%%%%%%%%%%%%%%%%%%%%%%
 
 \section*{APPENDIX A}
 In this appendix we shall study the first order perturbation theory of Hermitian operators. Consider an unperturbed Hermitian operator $A$, defined on a given Hilbert space, with orthonormal eigenvectors $|v_a\rangle$ and and corresponding eigenvalues $\lambda_a$ 
 \begin{equation}\label{A1}
 A |v_a\rangle = \lambda_a |v_a\rangle.
 \end{equation} 
 
 Suppose one adds a Hermitian perturbation $\delta A$ proportional to some small parameter $\epsilon$. The eigenvectors then become $|v_a\rangle + |\delta v_a\rangle$, with eigenvalues $\lambda_a + \delta\lambda_a$, where we assume $|\delta v_a\rangle$ and $\delta\lambda_a$ are given by power series in $\epsilon$ 
 \begin{eqnarray}\label{A2}
 |\delta v_a\rangle = |\delta_1 v_a\rangle +|\delta_2 v_a\rangle + \cdots \\ 
 \delta \lambda_a = \delta_1 \lambda_a +\delta_2 \lambda_a + \cdots \label{A3}.
 \end{eqnarray}
 Here $ |\delta_n v_a\rangle$ and $ \delta_n \lambda_a$ are presumably proportional to $\epsilon^n$. Substituting Eq. (\ref{A2}) and Eq. (\ref{A3}) into Eq. (\ref{A1}) and collecting the terms of the first order in $\epsilon$ and dropping the terms of higher order, we get : 
 \begin{equation}\label{A4}
 \delta A  | v_a\rangle + A  |\delta_1 v_a\rangle = \delta_1 \lambda_a | v_a\rangle + \lambda_a |\delta_1 v_a\rangle .
 \end{equation}
 
 To find $\delta_1 \lambda_a$, it is enough to take the inner product of Eq. (\ref{A4}) with $| v_a\rangle$. This gives
 
\begin{equation}\label{A5}
\delta_1 \lambda_a = \langle v_a|\delta A  | v_a\rangle .
\end{equation} 
 This tells us in the first order that the shifts in the eigenvalues are given by the expectation values of $\delta A$ in the unperturbed eigenvectors. But this argument has a bug in the case of degenerate Hermitian operators. To see what may go wrong in the degenerate case, we take the inner product of Eq. (\ref{A4}) with an arbitrary unperturbed eigenvector eigenvector $| v_b\rangle$. The result would be 
 \begin{equation}\label{A6}
 \langle v_b |\delta A  | v_a\rangle  = \delta_1 \lambda_a \delta_{ab} + (\lambda_a - \lambda_b) \langle v_b |\delta_1 v_a\rangle .
 \end{equation}
 For $a=b$, we get Eq. (\ref{A5}), but for $a\neq b$ we have
 \begin{equation}\label{A7}
 \langle v_b |\delta A  | v_a\rangle  = (\lambda_a - \lambda_b) \langle v_b |\delta_1 v_a\rangle\ \ \ \ \ a\neq b
 \end{equation}
 
 Now suppose there are two independent states $|v_a\rangle$ and $|v_b\rangle$ with the same eigenvalue. Apparently Eq. (\ref{A7}) is inconsistent for such states unless $\langle v_b |\delta A  | v_a\rangle$ vanishes, which of course does not need to be the case. To overcome this inconsistency, we have to choose the eigenvectors of the same eigenvalue in a special way. Suppose each eigenvalue $\lambda_a$ has a $N_a$-fold degeneracy with eigenvectors $| v_{a\mu}\rangle$ where $\mu=1, 2, \ldots, N_a$.  Because $\langle v_{b\mu} |\delta A  | v_{a\nu}\rangle$ form an Hermitian matrix, we can diagonalize it by using vectors $V_n = (V_{1n}, V_{2n}, \cdots, V_{N_a n})$, which are eigenvectors of this matrix with eigenvalues $\Delta_n$ i.e. 
 \begin{equation}\label{A8}
\sum_{\mu}\Big( \langle v_{b\mu} |\delta A  | v_{a\nu}\rangle\Big)  V_{\nu n} = \Delta_n V_{\mu n}.
 \end{equation}
Now, we can define eigenstates of $A$ with the same eigenvalue $\lambda_a$ as following
\begin{equation}\label{A9}
|u_{an}\rangle \equiv \sum_{\mu} V_{\mu n} |v_{a\mu}\rangle
\end{equation}
These  eigenvectors turn out to be normalized if one uses the orthonormality relation $\sum_{\mu} V^{*}_{\mu m} V_{\mu n}= \delta_{mn}$. It is easy to see that the eigenvectors $|u_{an}\rangle$ satisfy the following equation 

 \begin{equation}\label{A10}
 \langle u_{am} |\delta A  | u_{an}\rangle  = \Delta_n \delta_{mn},
 \end{equation}
 and hence if we use these eigenvectors we won't face with any inconsistency.\\

 %%%%%%%%%%%%%%%%%%%%%%%%%%
 \section*{APPENDIX B}
 
 In this appendix we shall briefly study the general solution of 
 \begin{equation}\label{B1}
 \dot{X}(t)= \mathcal{A} X(t),
 \end{equation}
 where $\cal{A}$ is a $d\times d$ matrix defined on a $d$-dimensional Hilbert space $\textbf{H}$ and $X(t)$ is a time-dependent vector with components $x_1(t), x_2(2), \ldots, x_d(t)$.  
 Let us label the eigenvalues of $\cal{A}$ as $\lambda^{(N_1)}_1, \lambda^{(N_2)}_2, \ldots, \lambda^{(N_k)}_k$, where $k$ is the number of eigenvalues and the indexes $N_i$ show $N_i$-fold degeneracies of eigenvalues. We say  $\cal{A}$ is a diagonalizable matrix if for each $\lambda^{(N_i)}_i$ there are $N_i$ linearly independent eigenvectors $Y^{(i)}_{\mu_i}$, corresponding to this eigenvalue i.e.
 \begin{equation}\label{B2}
{\cal{A}} Y^{(i)}_{\mu_i} = \lambda^{(N_i)}_i Y^{(i)}_{\mu_i},\ \ \ \ \ \mu_i= 1, 2, \ldots, N_i,
 \end{equation}
 where the eigenvectors $Y^{(i)}_{\mu_i}$ have been normalized i.e. $Y^{(i)\dagger}_{\mu_i} Y^{(j)}_{\nu_j}= \delta_{ij}\delta_{\mu_i, \nu_i}$. We note that for a diagonalizable matrix, with $k$ eigenvalues $\lambda^{(N_1)}_1, \lambda^{(N_2)}_2, \ldots, \lambda^{(N_k)}_k$,  we should have $N_1 +N_2 + \cdots N_k = d$. 
 
 If $\cal{A}$ in Eq. (\ref{B1}) is a diagonalizable matrix, once we find its eigenvectors and eigenvalues (according to Eq. (\ref{B2})), we immediately get the general solution as
 \begin{equation}\label{B3}
X(t)= \sum_{i=1}^{k} e^{t\lambda^{(N_i)}_i} \sum_{\mu_i=1}^{N_i} c^{(i)}_{\mu_i} Y^{(i)}_{\mu_i},
 \end{equation}
 where $c^{(i)}_{\mu_i}$ are integration constants and should be determined by the initial condition $X(0)$ :
 \begin{equation}\label{B4}
c^{(i)}_{\mu_i}=Y^{(i)\dagger}_{\mu_i} X(0).
 \end{equation}
 We say $\cal{A}$ is a non-diagonalizable matrix if at least there exists one defective eigenvalue i.e. there is an eigenvector $\lambda^{(N_j)}_j$, with $N_j$-fold degeneracy for which the number of independent eigenvectors is less than $N_j$. In such cases the solution of Eq. (\ref{B1}) is more complicated. Before we give the general solution of such cases, we need some new definitions. 
 
 For an arbitrary matrix $\cal{A}$, we say a vector $V\neq 0$ is a generalized eigenvector of rank $p\geqslant 1$, corresponding to eigenvalue $\lambda$, if 
 \begin{equation}\label{B5}
 {\Big({\cal{A}} -\lambda I\Big)}^p V=0\ \ \ \ \ and\ \ \ \ \   {\Big({\cal{A}} -\lambda I\Big)}^r V\neq 0\ \ \ \ for\ \ 0\leq r < p.
 \end{equation}
 It is clear that if $p=1$, then $V$ is an eigenvector which  we sometimes call \textit{trivial generalized eigenvector}. We define a length $p$-chain of generalized eigenvectors $V_i \neq 0$ (corresponding to eigenvalue $\lambda$) based on the eigenvector $V_1$, as a set $\Big\{ V_1, V2, \ldots, V_p\Big\}$, such that
 \begin{equation}\label{B7}
 {\Big({\cal{A}} -\lambda I\Big)} V_i=V_{i-1} \ \ i=2, 3, \ldots, p ;\ \ \ V_i = 0 \ \ for\ \ i>p .
 \end{equation}
 We note that in a $p$-chain we have ${\Big({\cal{A}} -\lambda I\Big)}^j V_j=0$ for each $V_j \in \Big\{ V_1, V2, \ldots, V_p\Big\} $.
 
 A fundamental theorem in linear algebra states that in a finite dimensional space, for a matrix $\cal{A}$ the number of independent generalized eigenvectors corresponding to an eigenvalue $\lambda$, with $N$-fold degeneracy, is equal to $N$.  Therefore,  all generalized eigenvectors of $\cal{A}$ establish a complete basis for the Hilbert space $\textbf{H}$. We note that $\cal{A}$ would be non-diagonalizable matrix if there exist nontrivial generalized eigenvectors.
 Before we write down the general solution of Eq. (\ref{B1}), let us look at a $N$-fold degenerate eigenvector $\lambda$. For example suppose this eigenvalue only has one trivial eigenvector $V_1$ and hence we can find a $N$-chain of generalized eigenvectors $V_1, V_2, \ldots, V_N$, corresponding to $\lambda$. Of course $X_1(t)= V_1 e^{\lambda t}$ is a solution. But, the generalized eigenvector $V_2$ can also gives us a new solution
 \begin{equation}\label{B8}
 X_2(t)= e^{\lambda t}\Big(V_2 + tV_1\Big).
 \end{equation}
 To check it, substitute $X_2(t)$ in Eq. (\ref{B1}) and use Eq. (\ref{B7}). Again, one can use the generalized eigenvector $V_3$ to find another independent solution 
 \begin{equation}\label{B9}
 X_3(t)=e^{\lambda t}\Big(\frac{t^2}{2!} V_1+ tV_2 + V3\Big).
 \end{equation}
 We may generalize this procedure to drive all solutions arising from the $N$-chain 
 \begin{equation}
 X_i(t)= e^{\lambda t}\Big(V_i + t V_{i-1} + \ldots +\frac{t^{i-1}}{(i-1)!} V_1\Big).
 \end{equation}
 
With this prescription, it would be easy to find the general solution of Eq. (\ref{B1}). Suppose $\lambda^{(N_i)}_i$ are eigenvalues of $\cal{A}$ with $r_i \leq N_i$ trivial generalized eigenvectors $Y^{(i)}_{\mu_i}={Y^{(i)}_{1}, Y^{(i)}_{2}, \ldots, Y^{(i)}_{r_i}}$. We show a $p^{(i)}_{\mu_i}$-chain of generalized eigenvectors, based on each eigenvector $Y^{(i)}_{\mu_i}$ as $\Big\{V^{(j)}_{\mu_j, a}\Big\}_{a=1}^{p^{(j)}_{\mu_j}}$, where $\sum_{\mu_j=1}^{r_j} p^{(j)}_{\mu_j}= N_j$ and $V^{(j)}_{\mu_j, 1} \equiv Y^{(j)}_{\mu_j}$. Based on what we stated above, the general solution of Eq. (\ref{B1}) has the following compact form 
\begin{equation}
X(t)=\sum_{i=1}^{k} e^{t\lambda^{(N_i)}_i} \sum_{\mu_i=1}^{r_i}\sum_{a=1}^{p^{(i)}_{\mu_i}} c^{(i)}_{\mu_i, a}\sum_{n=1}^{a} \frac{t^{n-1}}{(n-1)!} V^{(i)}_{\mu_i,a+1-n},
\end{equation} 
where $c^{(i)}_{\mu_i, a}$ are constants of integration, given by initial condition $X(0)$. We not that the generalized eigenvectors are chosen is such a way $V^{(i)\dagger}_{\mu_i,m}V^{(j)}_{\nu_j,n}=\delta_{ij}\delta_{mn}\delta_{\mu_i \nu_i}$ or, in other words, they are orthonormal vectors.\\

%%%%%%%%%%%%%%%%%%%%%%%%%%%%%
\section*{APPENDIX C}
%%%%%%%%%%%%%%%%%%%%

In this appendix we shall provide a short survey of the complete positivity concept. We define a positive operator $\cal{E}$ as a map from space of density matrices $M_d(C)$ to $M_d(C)$
\begin{equation}\label{C1}
\rho^{\prime}={\cal{E}}(\rho)\equiv  \sum_{\alpha=1}^{d^{2}}\lambda^{\alpha}{\bf E}^{\alpha}{\rho} {\bf E}^{\alpha\dagger},
\end{equation}
where $\lambda^{\alpha}$ are some real numbers, with the condition $\sum_{\alpha=1}^{d^2} \lambda^{\alpha}= d$, and ${\bf E}^{\alpha}$ are some orthonormal $d\times d$ matrices i.e. 
\begin{equation}\label{C2}
\sum_{i=1}^{d}\sum_{r=1}^{d}E_{ir}^{\alpha}E_{ir}^{\beta*}
=Tr\Big({\bf E}^{\alpha}{\bf E}^{\beta\dagger}\Big) =\delta^{\alpha\beta}.
\end{equation}
We say $\mathcal{E}$ is a completely positive map, if $\mathcal{E}\otimes I_d$ preserves the positivity on $M_d(C) \otimes M_d(C)$, where $I_d$ is the identity map on $M_d(C)$. In other words, $\mathcal{E}$ is completely positive if $\mathcal{E}\otimes I_d$ transforms a positive matrix in the larger space $M_d(C) \otimes M_d(C)$ to another positive one on that space. Now, there is a simple theorem of M. D. Choi \cite{Choi} which states that the positive map $\mathcal{E}$, defined in Eq. (\ref{C1}), would be a completely positive map if and only if all $\lambda^{\alpha}$ in Eq. (\ref{C1}) are non-negative real numbers. The proof is here. Let us define $\mathcal{R} \in M_d(C) \otimes M_d(C)$. The operator $\mathcal{E}\otimes I_d$ transforms this matrix to another matrix $\mathcal{R}^{\prime} \in M_d(C) \otimes M_d(C)$ 
\begin{equation}\label{C3}
{\cal R}'=\sum_{\alpha=1}^{d^{2}}\lambda^{\alpha}{\bf E}^{\alpha}{\cal R}{\bf E}^{\alpha\dagger},
\end{equation}
where ${\bf E}^{\alpha}$  act only on the first sector of $\mathcal{R}$. Now, we would like to define the vector $|w\rangle$ in the bigger Hilbert space $\textbf{H}_d\otimes \textbf{H}_d$ as
\begin{equation}\label{C4}
|w\rangle\equiv\sum_{m,n=1}^{d}D_{mn}|m\rangle|n\rangle,
\end{equation} 
where $\{|m\rangle\}_{m=1}^{d}$ is an orthonormal basis for $\textbf{H}_d$. Choose the matrix $\cal{R}$ as a pure density matrix with the following components 
\begin{equation}\label{C5}
{\cal R}\equiv\sum_{k,l,k',l'=1}^{d}C_{kl}C_{k'l'}^{*}|{k}\rangle|{l}\rangle\langle {k'}|\langle {l'}|,
\end{equation}
Calculating the expectation value of $\mathcal{R}^{\prime}$ with respect to $|w\rangle$, gives
\begin{eqnarray}\label{C6}
\langle w|{\cal  R}'|w\rangle&=&\sum_{\alpha=1}^{d^{2}}\lambda^{\alpha}\langle w|{\bf E}^{\alpha}{\cal R}{\bf E}^{\alpha\dagger}|w\rangle\nonumber\\
&=&\sum_{\alpha=1}^{d^{2}}\lambda^{\alpha}\sum_{1}^{d}D_{m'n'}^{*}D_{mn}C_{kl}C_{k'l'}^{*}E_{m'k}^{\alpha}E_{m k'}^{\alpha*}\delta_{n'l}\delta_{l'n}\nonumber\\
&=&\sum_{\alpha=1}^{d^{2}}\lambda^{\alpha}T\Big(Tr[{\bf C}{\bf D}^{\dagger} {\bf E}^{\alpha}]\Big){\Big(Tr[{\bf C}{\bf D}^{\dagger} {\bf E}^{\alpha}]\Big)}^{*},
\end{eqnarray}
where ${\bf D}=[D_{mn}]$ and ${\bf C}=[C_{mn}]$. We see if all $\lambda^{\alpha}$ are non-negative numbers, then $\langle w|{\cal  R}'|w\rangle \geq 0$. Because this is true for every pure density matrix, this would also be true for every density matrix. Therefore, the non-negativity of all $\lambda^{\alpha}$ gives the positivity of $\mathcal{E}\otimes I_d$ (and consequently the complete positivity of $\mathcal{E}$). On other hand, if we choose ${\bf D}{\bf C}^{\dagger} = {\bf E}^{\beta}$, for any arbitrary $\beta$, we shall find 
\begin{equation}\label{C7}
\langle w|{\cal  R}'|w\rangle = \sum_{\alpha=1}^{d^{2}}\lambda^{\alpha}T\Big(Tr[{\bf C}{\bf D}^{\dagger} {\bf E}^{\alpha}]\Big){\Big(Tr[{\bf C}{\bf D}^{\dagger} {\bf E}^{\alpha}]\Big)}^{*} = \lambda^{\beta},\ \ \ for \ 1\leq  \lambda^{\beta} \leq d^2
\end{equation}
where we have used the orthonormality condition Eq. (\ref{C2}). From Eq. (\ref{C7}), it is clear that if $\mathcal{E}\otimes I_d$ is a positive map ( or in other words $\langle w|{\cal  R}'|w\rangle \geq 0$ ), then we have $\lambda^{\alpha} \geq 0$ for all $\alpha$. \\

\end{document}